\documentclass{article}
\usepackage{mod_jheppub}
\usepackage{graphicx}
\usepackage{caption}
\usepackage{subcaption}
\usepackage{float}
\usepackage{holtpolt}
\usepackage{graphics,amstext,bbold,float}
\title{Ellipticity Weakens Chameleon Screening}
 \author{Clare Burrage,}
 \author{Edmund J. Copeland}
 \author{and James Stevenson}
 \affiliation{School of Physics and Astronomy, University of Nottingham, Nottingham, NG7 2RD, United Kingdom}

\emailAdd{clare.burrage@nottingham.ac.uk}
\emailAdd{ed.copeland@nottingham.ac.uk}
\emailAdd{james.stevenson@nottingham.ac.uk}

\date{today}
\abstract{The chameleon mechanism enables a long range fifth force to be screened in dense environments when non-trivial self interactions of the field cause its mass to increase with the local density.  To date, chameleon fifth forces have mainly been  studied for spherically symmetric sources, however the non-linear self interactions mean that the chameleon responds  to changes in the shape of the source differently  to gravity. In this work we focus on ellipsoidal departures from spherical symmetry and compute the full form of the chameleon force, comparing it's shape dependence to that of gravity. Enhancement of the chameleon force by up to 40\% is possible when deforming a sphere to an ellipsoid of the same mass, with an ellipticity $\simeq 0.99$.}

\begin{document}
\maketitle

%%%%%%%%%%%%%%%%%%%%%%

The cosmological constant problem is the most severe fine tuning problem in physics today.  It arises because matter fields are expected to contribute to the energy density of the vacuum at a scale many orders of magnitude higher than the value inferred from cosmological observations.  A convincing solution to this problem remains elusive, however we know that it will require us to go beyond the current standard models of cosmology and particle physics and that this solution must interact with both matter and gravitational fields.  A very common consequence of attempts to solve the cosmological constant problem is the introduction of a new, light scalar degree of freedom that couples to matter fields. Such a field is known as dark energy.  Whether this is a quintessence field,  a component of a massive graviton, arising from a string compactification \cite{Copeland:2006wr,Clifton:2011jh} or from another source entirely the presence of  light scalar degrees of freedom poses a problem.  Their coupling to matter means that they will mediate long range fifth forces that have not yet been detected on Earth or in the solar system \cite{Adelberger:2003zx}.  This tension between theory and observation is resolved by the introduction of screening mechanisms, which allow the properties of the  field, and the force that it mediates,  to vary depending on the environment \cite{Khoury:2013tda} at the cost of making the scalar field theory non-linear.  \\
\\
A number of attractive scalar theories admitting screening behaviour have been constructed to date.  Amongst the more commonly studied are the Chameleon, Dilaton and the Symmetron models \cite{JK2,Brax:2010gi,Hinterbichler:2010es} where non-linearities in the scalar potential or form of the coupling to matter result in the mass of the field, or the strength of the coupling becoming dependent on the environment. For suitable parameter choices this allows the scalar force to be  suppressed in regions of higher density including those used in experimental searches for fifth forces  making them, in principle, compatible with experimental constraints.  Despite being explicitly designed to evade the constraints of current fifth force searches, a  benefit of these theories is that alternative scenarios can be devised to search for the existence of  such scalar fields in laboratory based experiments.  In a laboratory vacuum the low density ensures that sufficiently small objects are not screened from the scalar field and are thus sensitive probes of dark energy. In this regime the force sourced by the dark energy scalar  could  significantly exceed the gravitational interaction, and yet no deviation from general relativity would be seen with larger sources or in less diffuse environments.  A number of ways of probing the chameleon screening mechanism in the laboratory have been proposed or implemented in recent years.  In \cite{TJ,LNG,BR}, neutron spectroscopy experiments have been conducted at an energy scale of $10^{-14}$ eV, that constrain departures from Newtonian gravity over micron distances and can be used to  introduce new bounds on the parameter space of the chameleon model. Casimir-like experiments are also under way to study the chameleon field over very short distance scales \cite{Brax:2007vm,Brax:2010xx}.   In the near future, much of the remaining parameter space is set to be probed with the development of ultra-cold atom interferometry experiments \cite{CL} that,  in principle, can detect even a chameleon field with Planck suppressed couplings. \\
\\
In screened scalar field theories governed by the presence of non-linearities, it is natural to question whether the shape of the source object changes the efficiency of screening. Source shape dependence has been considered in reference \cite{VSD} for theories exhibiting Vainshtein screening,  where the comparison of several sources (spherical, cylindrical and planar)  showed a  reduction in the  effectiveness of the screening for the non-spherical geometries. In a cosmological setting, this  motivates targeting searches for the Galileon field at matter morphologies such as walls or filaments.\footnote{In contrast to the models previously mentioned, Vainshtein screening occurs in theories with non-linearities present in the kinetic terms.}  The purpose of this work is to examine whether deforming the source object from a sphere to an ellipsoid generates any shape enhancements that heighten the detectability of a  fifth force screened by the chameleon mechanism. Due to the complexity of the calculations, our analysis will be restricted exclusively to the chameleon model although,  as dilaton and symmetron models have very similar phenomenology to the chameleon for spherical sources, we expect  that the sensitivity of the chameleon to the shape of the source object  derived in this article  will extend to the dilaton and symmetron  models as well. A related study has previously been performed for the chameleon in \cite{EA} which exploits analogies between the chameleon field equations and equations in electrostatics  to analyse the chameleon field around an ellipsoidal source. It should be noted however, that only the far field solution was presented in reference \cite{EA} and that this does not hold in the  regime of interest in this paper, where a more complete treatment is required and presented. \\
\\
We will first lay out the theoretical framework for the chameleon in Section \ref{The_Chameleon_Theory}.  We will then review the form of the gravitational field profile around an ellipsoidal source in Section \ref{Gravity_Around_Ellipse}.  Section \ref{Mapping_the_Chameleon_Profile} then contains the derivation of the chameleon field profile around an ellipsoidal source.  We conclude in Section \ref{Comparisons_to_Gravity} with a discussion of the implications of these results for the possibility of detecting the chameleon in a laboratory experiment.  In the appendices further details are given  justifying various approximations made in our calculations.
\section{The Chameleon Theory}\label{The_Chameleon_Theory}
The chameleon field is described by the following action:
\begin{equation}
\mathcal{S} = \int d^4x \sqrt{-g} \left\{\frac{M_{Pl}^2}{2}R - \frac{1}{2}\partial_\mu\varphi\partial^\mu\varphi - V(\varphi)\right\}  + \int d^4x \>\mathcal{L}_m \big(\psi_{m}, \Omega^{2}(\varphi) g_{\mu\nu}\big) \;,
\label{eq:action}
\end{equation}
where the chameleon field $\varphi$ is minimally coupled to gravity and non-minimally coupled to matter. Here $M_{Pl}^2 =1/(8\pi G)$ is the reduced Planck mass and $V(\varphi)$ is the scalar potential.  This simplified model assumes a universal minimal coupling between the conformally rescaled metric $\tilde{g}_{\mu\nu} = \Omega^{2}g_{\mu\nu}$ and each matter species, $\psi_m$. 
We are able to linearise the function describing the  conformal factor  $\Omega = 1 + \frac{\varphi}{M}$, where $M$ denotes the coupling strength between the chameleon and matter, as we will always work in the regime $\varphi \ll M$. Conformal couplings of this kind appear for  string theory  dilaton fields, and also when $f(R)$ theories of modified gravity are re-written in the Einstein frame \cite{Fujii:2003pa,Brax:2008hh}.   The density dependent behaviour of the chameleon requires that the chameleon potential $V(\varphi)$ contains non-trivial self interaction terms.  
A common choice that we will also make in this work is $V(\varphi) = \frac{\Lambda^5}{\varphi}$ although other exponents of the field have been considered \cite{Mota:2006fz}.  \\
\\
When the matter source is  static and non-relativistic the chameleon field equation that results from the action in Equation (\ref{eq:action}) is:
\begin{equation}
\nabla^2\varphi = \frac{\partial V}{\partial\varphi} + \rho\frac{\partial\Omega}{\partial\varphi}\;,
\end{equation}
where $\rho$ is the matter energy density. This indicates that the chameleon can be considered as a field moving in an effective potential:
\begin{equation}
V_{\rm eff} = V(\varphi) + \rho \Omega(\varphi)\;,
\end{equation}
when $\rho$ is assumed to be constant. 
The mass of the field follows by considering  fluctuations around the vacuum state $\varphi_{\rm min}$, such that $V_{\rm eff}^{\prime}(\varphi_{\rm min})=0$. 
\begin{equation}
m^2 = \frac{\partial^2{V_eff}}{\partial{\varphi^2}}  = \frac{\partial^2 V}{\partial{\varphi^2}} + \rho\frac{\partial^2\Omega}{\partial\varphi^2}\;.
\end{equation}
Substituting the choice of potential $V(\varphi)$ and  conformal factor $\Omega(\varphi)$ described above, we find that the position of the minimum of the effective potential and the mass of the chameleon field are:
\begin{equation}
\varphi_{\rm min} = \Big(\frac{\Lambda^5 M}{\rho}\Big)^{1/2}\;, \>\>\>\>\> m^2 = 2\Big(\frac{\rho^3}{\Lambda^5M^3}\Big)^{1/2}\;,
\end{equation}
where it is evident that in regions of higher density the field value is smaller and the force has  a  shorter interaction range, due to a larger mass of the field  and correspondingly reduced Compton wavelength. \\
\\
The chameleon theory as introduced here does not solve the cosmological constant problem, a fine tuned constant term is required in the scalar potential if the chameleon is to drive the acceleration of the expansion of the universe.   Reconciling the chameleon with cosmological observations requires the mass of the  field in the cosmological vacuum today to be  of the order of the Hubble scale $m \simeq H_0$ \cite{Brax:2011aw,Wang:2012kj}.  The chameleon model relies on the presence of non-linear terms in its potential, and therefore when considered as a quantum effective field theory  non-renormalisible operators are inevitable.  The theory is not protected from quantum corrections, and the scalar potential must be considered fine tuned, and therefore the theory should  only be treated  as a low energy effective theory.  These quantum corrections can become important in laboratory searches, and in the early universe \cite{Erickcek:2013oma,Erickcek:2013dea,Upadhye:2012vh}.  We acknowledge that this fine tuning is a potential issue for chameleon theories but to attempt to solve this problem is beyond the scope of this work. Moreover this work is primarily aimed at demonstrating the more general point that density dependent couplings of scalar fields could well be detected within laboratory experiments, and that the strength of the force is sensitive to the shape of the source.

\section{Gravity around an Ellipsoidal Source}\label{Gravity_Around_Ellipse}
Before turning to the chameleon, we will first review the Newtonian gravitational field profile surrounding an ellipsoidal source. 
Our analysis of ellipsoidal geometries uses  spherical prolate coordinates.  These are related to Cartesian co-ordinates via the following transformations
\begin{equation}
\begin{split}
& x = a\sqrt{(\xi^2 - 1)(1 - \eta^2)}\> \cos(\phi)\;, \\
& y = a\sqrt{(\xi^2 - 1)(1 - \eta^2)}\> \sin(\phi) \;,\\
& z = a\xi\eta\;,
\end{split}
\end{equation} 
and are defined such that the major axis of the ellipse lies along the $z$-direction with foci at $ z = \pm a $.
The co-ordinate $\phi$ represents the conventional azimuthal angle appearing in spherical polar co-ordinates whereas $\xi$ and $\eta$ are analogous to the radial and polar angular components respectively. The coordinate ranges are $ +1 < \xi < \infty$, $ -1 < \eta < +1$ and $ 0 < \phi < 2\pi$. The only dimensionful parameter entering the coordinate system is the focal length $a$ which controls the size  of the ellipsoid.  The gravitational potential, $\Phi$, is governed by Poisson's equation, 
\begin{equation}
\nabla^2 \Phi = -\frac{\rho}{2M_{Pl}^2}\;.
\end{equation}
To determine the gravitational field profile around a compact ellipsoidal source of constant density, $\rho_{\rm obj}$, contained within $1\leq \xi\leq \xi_0$, and embedded in a zero density background we need to solve the following system of equations:
\begin{align}\label{Laplace-Gravity}
\nabla^2 \Phi &= 0\;,  &\text{ for }  \xi_0 \leq \xi < \infty \;,\\
\nabla^2 \Phi &= -\frac{\rho_{\rm obj}}{2M_{Pl}^2}\;,  &\text{ for }  1 \leq \xi \leq \xi_0\;.
\end{align}
The shape parameter describing the source $\xi_0$ may be related to other common ellipsoidal parametrisations such as the ellipticity $e$ and the compression factor $\zeta$ via the following relations
\begin{equation}
e = \frac{1}{\xi_0}, \hspace{1 cm} \zeta = \frac{\sqrt{\xi_0^2-1}}{\xi_0}\;,
\end{equation}
where a smaller value of $\xi_0$ reflects a lower compression factor or equivalently a higher ellipticity. We impose boundary conditions to ensure that $\Phi$ and $\nabla\Phi$ are continuous at $\xi=\xi_0$, that $\Phi$ decays as $\xi \rightarrow \infty$ and is regular at $\xi=1$. This problem has a separable solution which we write as $ \Phi = \tilde{\Phi}(\phi)\sum_{l=0}^{\infty}X_l(\xi)H_l(\eta) $, where we have written the $\eta$ and $\xi$ dependent factors as a sum over an orthonormal set of basis functions. As the system under consideration has azimuthal symmetry, we are able to make the simplifying  assumption that $\Phi$ is independent of $\phi$, and thus $\tilde{\Phi}$ is constant.  Outside the source, $\xi\geq \xi_0$, the Laplace equation 
\begin{equation}
         \frac{(\xi^2 - 1)(1 - \eta^2)}{(\xi^2  - \eta^2)}\left\{\frac{1}{X_l(\xi)}\frac{\partial}{\partial\xi}(\xi^2 - 1)\frac{\partial X_l(\xi)}{\partial\xi} + \frac{1}{H_l(\eta)}\frac{\partial}{\partial\eta}(1 - \eta^2)\frac{\partial H_l(\eta)}{\partial\eta} \right\} = 0\;,
\end{equation}
can be separated into two copies of Legendre's equation
\begin{subequations}
\begin{align}
\frac{\partial}{\partial\xi}(\xi^2 - 1)\frac{\partial X_l(\xi)}{\partial\xi} -  \lambda_l X_l(\xi) & = 0\;, \label{xi-equation}\\
\frac{\partial}{\partial\eta}(1 - \eta^2)\frac{\partial H_l(\eta)}{\partial\eta} +  \lambda_l H_l(\eta) & = 0\;, \label{eta-equation} 
\end{align}
\end{subequations}
where $\lambda_l$ is a separation constant. As we show in equation (\ref{lambdal-massless}) of Appendix \ref{A2}, this  leads to the specific relation $\lambda_l=l(l+1)$ for any  non-negative integer $l$. This has the effect of reducing these equations to Legendre form. The corresponding Legendre functions of the first kind $P_l(x)$ are defined on the interval $(-\infty, \infty)$, whereas those of the second kind $Q_l(x)$ are defined on the interval $(1, \infty]$.  It means therefore that solutions to the $\xi$ equation above can be Legendre functions of either the first or second kind, whereas solutions to the $\eta$ equation can only be Legendre functions of the first kind. The full solution for $\Phi$ in the exterior of the source is a linear combination of these Legendre functions:
\begin{equation}
\Phi = \sum_{l =0}^{\infty} A_l P_l(\eta)P_l(\xi) + B_l P_l(\eta) Q_l(\xi)\;,\mbox{ for } \xi \geq \xi_0\;.
\label{eq:Phihomo}
\end{equation}
where $A_l$ and $B_l$ are $l$-dependant constants.
The first term in this expression diverges as $\xi\rightarrow \infty$, and so we set $A_l=0$ in order to ensure that the field decays at infinity.  It remains to determine the form of $\Phi$ in the interior of the source.  This can be written as the sum of  a homogeneous solution of the form given in Equation (\ref{eq:Phihomo}) and a particular integral, $\Phi_{\rm PI}$ that takes the form:
\begin{equation}
\Phi_{\rm PI} = -\frac{\rho_{\rm obj} a^2}{8M_{Pl}^2} \big(\xi^2 - 1\big)\big(1 - \eta^2\big) \equiv - \frac{\rho_{\rm obj} a^2}{12M_{Pl}^2} (\xi^2 - 1)\big[P_0(\eta) - P_2(\eta)\big]\;.
\end{equation}
Imposing that the solution does not diverge at the centre of the source, $\xi=1$, we can write the full interior solution as 
\begin{equation}\label{Interior_Ini}
\Phi =  \sum_{l=0}^{\infty} C_l P_l(\eta) P_l(\xi)     - \frac{\rho_{\rm obj} a^2}{12M_{Pl}^2} (\xi^2 - 1)\big[P_0(\eta) - P_2(\eta)\big]\;, \hspace{1 cm} 1<\xi \leq \xi_0\;.
\end{equation}   
The surviving unknowns $B_l$ and $C_l$ are identified by imposing the remaining boundary conditions that ensure that the field and its first derivatives are continuous across the surface of the object at $\xi_0$. This leads to the final solution for the form of the gravitational field profile for an ellipsoidal source:
\begin{equation}
\Phi =\frac{M_{\rm obj}}{8\pi M_{Pl}^2a}\left\{\begin{array}{lc}
  Q_0(\xi) - P_2(\eta)Q_2(\xi)\;, &\xi>\xi_0\;.\\
Q_0(\xi_0)[1-P_2(\eta)P_2(\xi)]  & 1<\xi \leq \xi_0\;.\\
+\frac{1}{\xi_0(2P_2(\xi_0)+1)}\left\{P_2(\xi_0)-P_2(\xi)+2P_2(\eta)P_2(\xi)+2P_2(\eta)P_2(\xi)[1+P_2(\xi_0)]\right\} & 
\end{array}\right.
\end{equation}
where $M_{\rm obj}$ is the mass of the source. 

\section{The Chameleon Profile Around an Ellipsoidal Source}\label{Mapping_the_Chameleon_Profile}

\begin{figure}[ht]
\centering
\includegraphics[width=\textwidth, trim = 1.7cm 19cm 1cm 2cm]{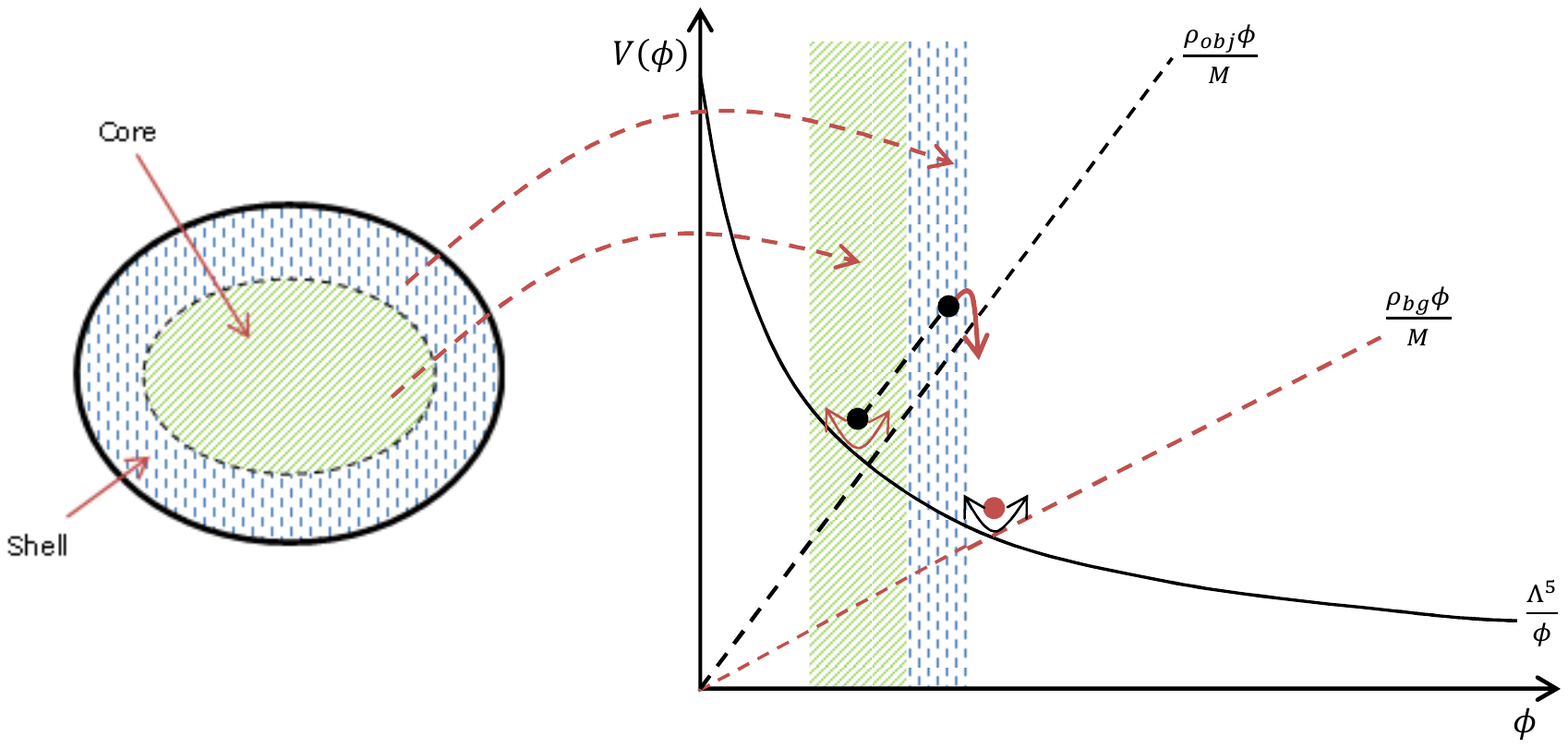}
\caption{The chameleon field profile decomposes into  core, shell and exterior regions. The relevant regime within the potential is indicated for each of the three regions.}\label{Shell_Decomposition}
\end{figure}

To find the chameleon field profile around an ellipsoidal source, we follow a very similar procedure to that taken to solve the gravitational equations in the previous section.  The differences arise when we are considering the form of the chameleon field profile in the interior of the source. As illustrated in figure \ref{Shell_Decomposition}, we expect that an ellipsoidal source will give rise to similar chameleon behaviour to that due to a spherical object, meaning that a sufficiently large source will form a `thin-shell'.  This occurs when the chameleon field reaches the field value that minimises the effective potential in the interior of the source.  Due to the large mass of fluctuations around this value, the field is unable to move from this minimum and remains stuck at this value for almost the whole of the interior of the object, it only changes its value in a thin-shell near the surface \cite{JK2}.  We expect the same behaviour for an ellipsoidal source, with the difference that the thin shell is defined by two ellipsoids.  This behaviour of the solution is illustrated in figure \ref{Shell_Decomposition}.
As for spherical objects a core region where the field reaches the minimum of the effective potential may not exist.  If it is absent the object is referred to as `thick-shell' and the field profile just takes the form of a standard Yukawa force law. If the mass of the field is negligible this is exactly analogous to the  gravitational potential (with the possibility of a different coupling constant) determined in the previous section, because the field never reaches the non-linear region of the  chameleon potential. 
Therefore we will solve the field equations assuming that a core region exists, and then determine the properties of the source that allow this solution to exist. We break down the chameleon field equations in the following way:
\begin{equation}
\begin{split}
\nabla^2\varphi = m_{\rm bg}^2(\varphi - \varphi_{\rm min}^{\rm (bg)})\;, \hspace{2 cm} \xi_0 \leq &\xi < \infty\;, \\
\\
\nabla^2\varphi = \frac{\rho_{\rm obj}}{M}\;, \hspace{2 cm} \xi_{\rm core} \leq &\xi \leq \xi_0\;, \\
\\
\nabla^2\varphi = m_{\rm obj}^2(\varphi - \varphi_{\rm min}^{\rm (obj)})\;, \hspace{2 cm} 1 \leq &\xi \leq\xi_{\rm core}\;,
\end{split}
\label{eq:chamethin}
\end{equation}
where $\varphi_{\rm min}^{(\rm bg)}$ and  $m_{\rm bg}$ are the field value and mass at the minimum of the effective potential in the background of the experiment, where the density is assumed to be $\rho_{\rm bg}$, whilst $\varphi_{\rm min}^{(\rm obj)}$ and  $m_{\rm obj}$ are the equivalent values in the interior of the source, which has density $\rho_{\rm obj}$.\\
\\
In what follows we will find that we can always neglect the mass of the chameleon as long as we are only interested in studying the behaviour of the field close to the source. At distances far from the source  the Yukawa suppression  becomes important and the mass of the field plays an important role in determining when this occurs.  Closer to the source the important dimensionless (in natural units) combination of parameters is $ma$; the mass of the chameleon multiplied by the focal length of the ellipse.  Within the available chameleon parameter space and considering values for the size and mass of the source object suitable for laboratory experiments the combination $ma$ is always small, and therefore the mass of the chameleon can be safely neglected in the analysis that follows.  In Appendix \ref{A2} we explore this approximation in more detail and compute the first order corrections in $ma$ to the massless solution, this follows previous work deriving ellipsoidal solutions to the Helmholtz equation in \cite{MF1,MF2,SBP,CB}.\\
\\
Neglecting the masses of the chameleon in equations (\ref{eq:chamethin}) reduces them to the forms of the Laplace and Poisson equations,  solved for the gravitational case in the previous section. The chameleon thin-shell solution only differs from the gravitational solution by the existence of the core region.  We still impose that the field must decay at infinity, and be regular at the centre of the ellipsoid, $\xi=1$, and that the field and its first derivative are continuous at the surface of the source and at the surface of the core region, $\xi_{\rm core}$.  The position of the surface of the core will be determined from these boundary conditions.  
Therefore the solution has the form:
\begin{align}\label{Solutions_Set_General}
 \varphi = & \sum_{l =0}^{\infty} V_l P_l(\eta)P_l(\xi) + \varphi_{\rm min}^{\rm (obj)}\;, & 1 \leq \xi \leq \xi_{\rm core}\;,\\
 \varphi = & \sum_{l=0}^{\infty} \Big(C_l P_l(\eta) P_l(\xi) + D_l P_l(\eta) Q_l(\xi)\bigg) - \frac{1}{6}\frac{\rho_{\rm obj}}{M}(\xi^2 - 1)\Big[P_0(\eta) - P_2(\eta)\Big]\;, & \xi_{\rm core} \leq \xi \leq \xi_0\;,\\
 \varphi =& \sum_{l=0}^{\infty} W_l P_l(\eta) Q_l(\xi) + \varphi_{\rm min}^{\rm (bg)}\;,
 &\xi_0 \leq \xi \leq \infty\;.
\end{align}
To determine the remaining unknowns, $V_l$, $C_l$, $D_l$, $W_l$\footnote{Note that the values of the constants $C_l$ do not take the same values as the constants found in the gravitational case in the previous section.}  and $\xi_{\rm core}$ we impose continuity of $\varphi$, and $\partial \varphi/\partial \xi$ across the two surfaces at $\xi_0$ and $\xi_{\rm core}$.  Importantly, these boundary conditions must apply for all $\eta$, thus the boundary conditions are imposed separately for each $P_l$.  For example continuity of $\varphi$ at $\xi_{\rm core}$ requires
\begin{subequations}
\begin{alignat}{3}
\varphi_{\rm min}^{\rm (obj)} &= C_0  + D_0 Q_0(\xi_{\rm core}) - \frac{1}{6}\frac{\rho_{\rm obj}}{M}(\xi_{\rm core}^2 - 1)\;, \\[ 10 pt]
V_2 P_2(\xi_{\rm core}) &= C_2 P_2(\xi_{\rm core}) + D_2 Q_2(\xi_{\rm core}) + \frac{1}{6}\frac{\rho_{\rm obj}}{M}(\xi_{\rm core}^2 - 1)\;, \\[ 14 pt]
 V_l P_l(\xi_{\rm core}) &=  C_l P_l(\xi_{\rm core}) + D_l Q_l(\xi_{\rm core})\;, \hspace{2 cm} l = 1,3,4,5,... \label{boundary_example} \vspace{1 cm} 
\end{alignat}
\end{subequations}
After applying all of the above boundary conditions the final form of the chameleon field (to all orders in $l$) in the exterior of an ellipsoidal source is
\begin{equation}\label{exterior-soln-varphi}
\varphi = \frac{1}{2}\frac{\rho_{\rm obj} a^2}{3M}\>\Big\{\xi_0(\xi_0^2 - 1) - \xi_{\rm core}(\xi_{\rm core}^2 - 1)\Big\}\bigg(Q_0(\xi) - P_2(\eta) Q_2(\xi)\bigg)\;, \mbox{ for } \xi_0 \leq \xi < \infty\;,
\end{equation}
and the position of the surface of the core, $\xi_{\rm core}$, is given implicitly by the expression:
\begin{equation}\label{Shell_equation}
\frac{6}{\rho_{\rm obj} a^2}\bigg(\frac{M^3 \Lambda^5}{\rho_{\rm bg}}\bigg)^{1/2} + (\xi_{\rm core}^2 - 1)\bigg\{1 +2\xi_{\rm core}Q_0(\xi_{\rm core})\bigg\} = (\xi_0^2 - 1)\bigg\{1 + 2\xi_0 Q_0(\xi_0)\bigg\}\;.
\end{equation}
These results differ from  the results presented in \cite{EA} due to the appearance of the quadrupole term $P_2(\eta)$ in equation (\ref{exterior-soln-varphi}), which is included in our analysis and was neglected in previous work.  Therefore we do not find the enhancement of the field at the `tips' of the ellipsoid that was suggested in \cite{EA}.  The thin shell solution holds whenever there is a solution to equation (\ref{Shell_equation}) which gives a real and strictly positive value for $\xi_{\rm core}$.  This is difficult to determine analytically, but is straightforward to do numerically, and we will investigate these solutions further in the following section.
For a sufficiently thin shell ($\xi_0 - \xi_{\rm core} \ll 1$), an analytic expression for the position of the interior surface of the shell can be found by manipulating equation (\ref{Shell_equation}):
\begin{equation}
\xi_0 - \xi_{\rm core} = \frac{6}{\rho_{\rm obj} a^2} \bigg(\frac{M^3 \Lambda^5}{\rho_{\rm bg}}\bigg)^{\frac{1}{2}}\frac{1}{(3\xi_0^2 - 1)Q_0(\xi_0)}\;.
\label{eq:lincore}
\end{equation}
As a consistency check of our result we check that this expression recovers the familiar expression for the thin shell of a sphere when we take the spherical limit $\xi_0\rightarrow \infty$, with $a$ fixed.  For large $\xi$ the Legendre function of the second kind becomes $Q_0(\xi) = 1/2\xi$, to first order. For objects that retain a thin shell as they are deformed from an ellipsoid to a sphere we assume that both $\xi_0$ and the shell position $\xi_{\rm core}$ are large. Equation (\ref{eq:lincore}) then becomes
\begin{equation}
\frac{6 M }{3\rho_{\rm obj} a^2}\varphi_{\rm min}^{\rm (obj)}  = \xi_0^2 - \xi_{\rm core}^2\;,
\end{equation} 
where we have assumed  $\varphi_{\rm min}^{\rm (obj)}\ll\varphi_{\rm min}^{\rm (bg)}$, which is valid as long as $\rho_{\rm obj}\gg \rho_{\rm bg}$.  At large $\xi$ the ellipsoidal coordinate $\xi$ can be related to the radial coordinate in spherical polars by  $\xi = r / a$ (this is derived in Appendix \ref{A3}).  Hence, identifying $\xi_{\rm core}^2 = S^2/a^2$ and $\xi_0^2 = R^2/a^2$ leads to 
\begin{equation}
S^2 = R^2 - \frac{2M}{\rho_{\rm obj}}\varphi_{\rm min}^{\rm (obj)}\;,
\end{equation}
which is the familiar expression for the thin shell of a sphere first derived in  \cite{JK2}.

\section{Comparisons to Gravity}\label{Comparisons_to_Gravity}
\begin{figure}[ht]
\begin{center}
\centerline{\scalebox{0.65}{\includegraphics[trim={1cm 0.5cm 0 0.5cm}]{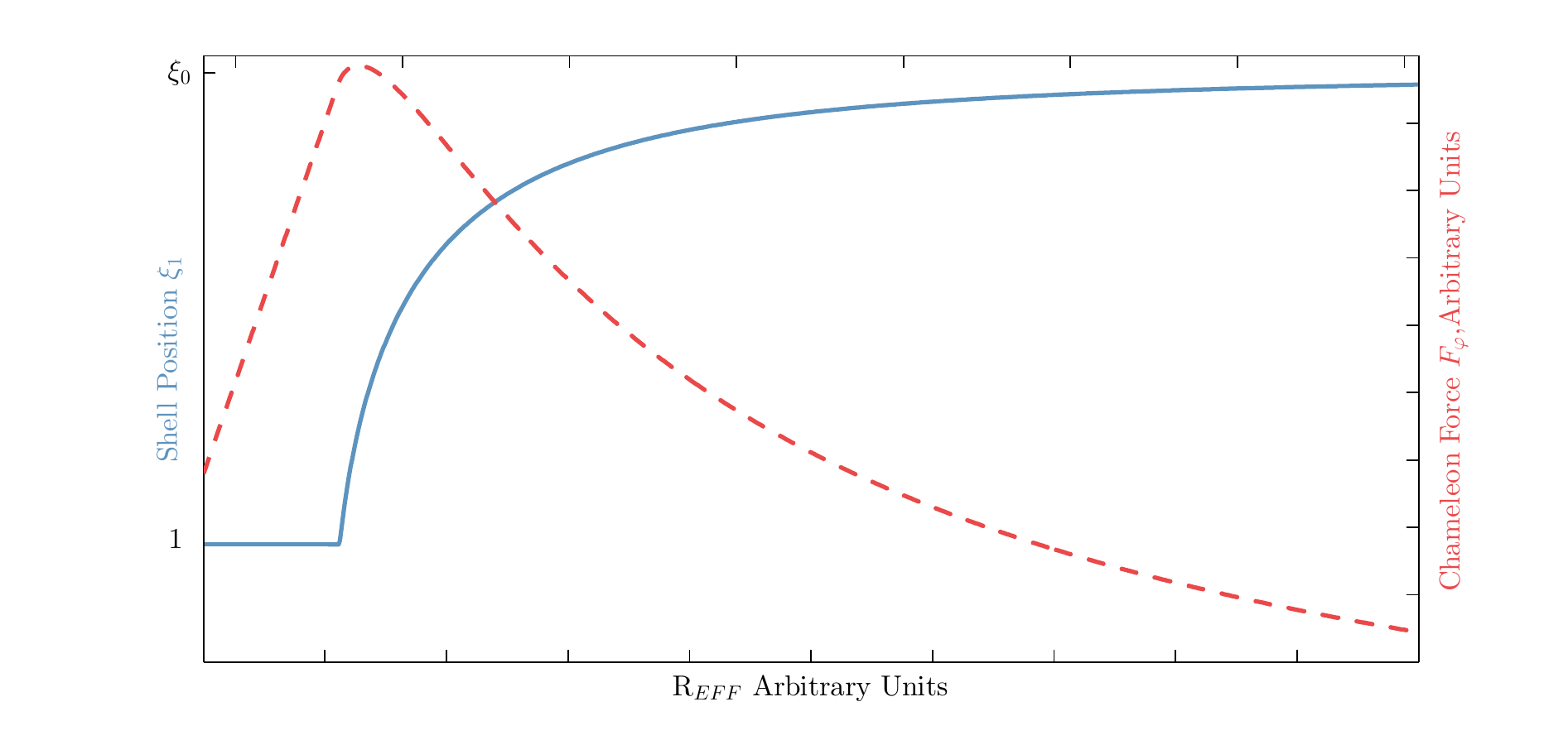}}}  
\caption{The blue line represents the scaling of the relative size of the shell region within an ellipsoidal source of size $R_\textrm{EFF}$. At the far left, a constant value is approached as the shell solution pervades the entire object and the core region is absent. As the effective radius is increased, the extent of the shell region begins to diminish due to the emergence of a core region. The dotted red line shows the corresponding  dependence of the magnitude of the chameleon force sourced by the compact object.  The  maximum force magnitude occurs when the radius of the source corresponds to the formation of a core region.}\label{fig:Thin_Shell_Effect}
\end{center}
\end{figure}

\begin{figure}[ht]
\begin{center}
\centerline{\scalebox{0.65}{\includegraphics{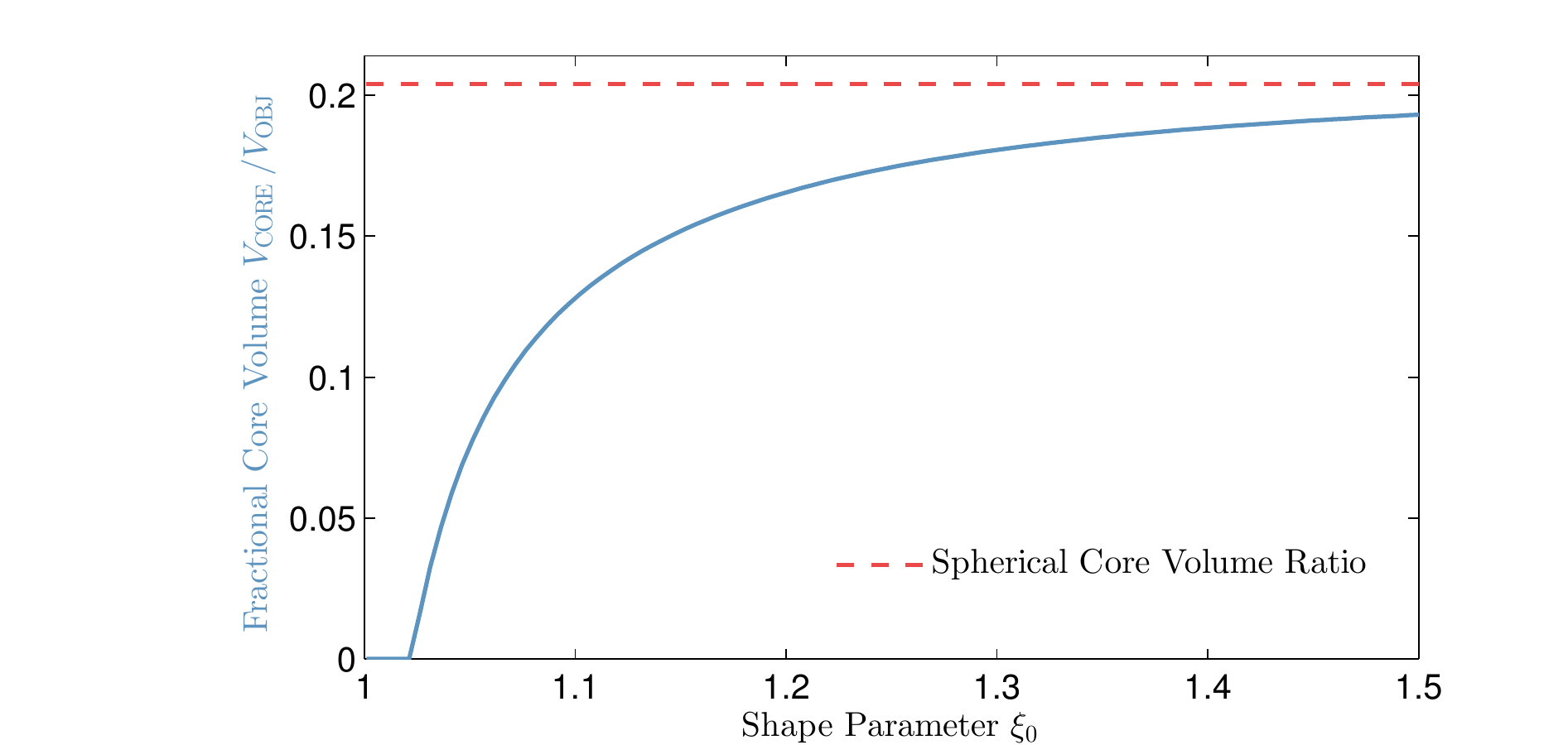}}}  
\caption{Example of how the size of the core region decays as the ellipticity of the source object increases (lower $\xi_0$). The line asymptotes to the core volume for the spherical scenario as $\xi_0 \rightarrow \infty$, represented by the red dashed line. This figure was generated for an object of size $R_{\rm{EFF}} = 4$m as this value captures the whole range of scaling effects. $M$ and $\Lambda$ were set as $M_{\rm{PL}}$ and $10^{-12}$ GeV respectively.}\label{fig:Core_Scaling}
\end{center}
\end{figure}

\begin{figure}[ht]
\begin{center}
\centerline{\scalebox{0.6}{\includegraphics{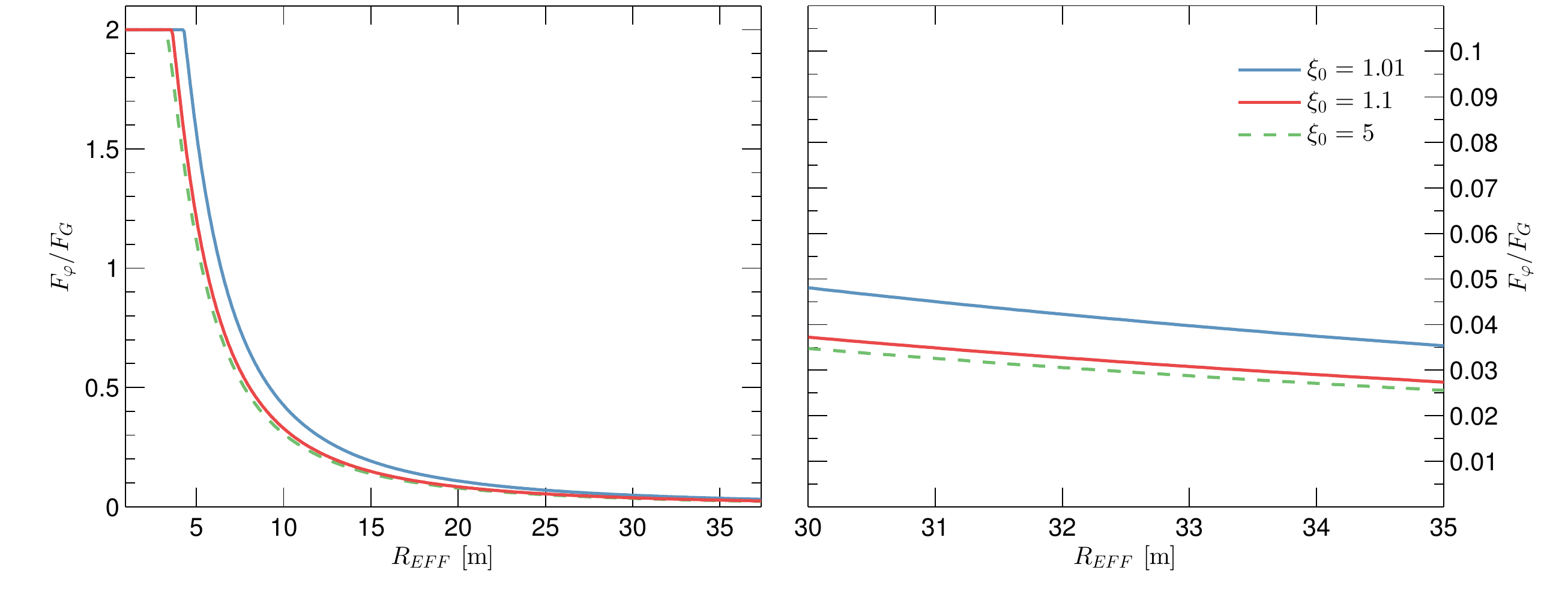}}}  
\caption{The ratio of the chameleon to gravitational forces as a function of the size of the source mass, parameterised by the radius of a sphere of equal mass $R_{\rm EFF}$.  The different lines show sources of different ellipticities, with the green line being the most spherical and the blue line the most ellipsoidal.  The right hand plot is a magnification of a region of the left hand plot to show that the shape enhancement of the chameleon force persists as the size of the source is increased.  To obtain numerical values the parameters were chosen to be $\rho_{\rm obj} = 10^3$, $\rho_{\rm bg} = 10^{-14}$ kgm$^{-3}$, $M = M_{Pl}$ and $\Lambda$ set at the dark energy scale $10^{-12}$ GeV
}\label{Sphere_Ellipse_Comparison}
\end{center}
\end{figure}

Our intention is to study if ellipsoidal deformations impact on the efficiency of chameleon screening and as a result on the detectability of the chameleon. The chameleon force on a test particle due to the ellipsoidal source can be found from the chameleon potential derived in the previous section as $\vec{F}_\varphi = \vec{\nabla}\varphi /M$. We have found that outside the source, $\xi > \xi_0$ the chameleon and gravitational  potentials obey the same functional form, and therefore the ratio of the corresponding forces is a constant independent of the coordinates:
\begin{equation}
\frac{F_\varphi}{F_G} = 2\bigg(\frac{M_{PL}}{M}\bigg)^2\bigg\{1 - \frac{\xi_{\rm core}(\xi_{\rm core}^2 - 1)}{\xi_0(\xi_0^2 - 1)}\bigg\}\;.
\end{equation}
As expected the chameleon force is maximised when there is no core region in the interior, $\xi_{\rm core}=1$, and in this case if $M$ is chosen to lie at the Planck scale, the chameleon force is twice the gravitational force. Once a core region develops the chameleon force is suppressed, this is the essence of chameleon screening, and this behaviour can be clearly seen in figure \ref{fig:Thin_Shell_Effect}.  
To determine how changing the shape of a source affects the screening we fix the mass of the object, and only deform its shape.  As we will assume a fixed uniform density for the interior of both spheres and ellipsoids we will refer to the size of any given ellipsoid in terms of the parameter $R_{\rm EFF}$ which is the radius of a sphere of the same mass, $R_{\rm EFF} = a (\xi_0(\xi_0^2 - 1))^{1/3}$, this expression is derived in Appendix \ref{A3}, equation (\ref{eq:reff}). Unless noted otherwise, any figures quoted are obtained by taking the source object and background densities to be $1$ gcm$^{-3}$ and $10^{-17}$ gcm$^{-3}$ respectively, which are appropriate values for laboratory experiments.
\begin{figure}[ht]
\centering
\begin{subfigure}{\textwidth}
\includegraphics[width=\textwidth]{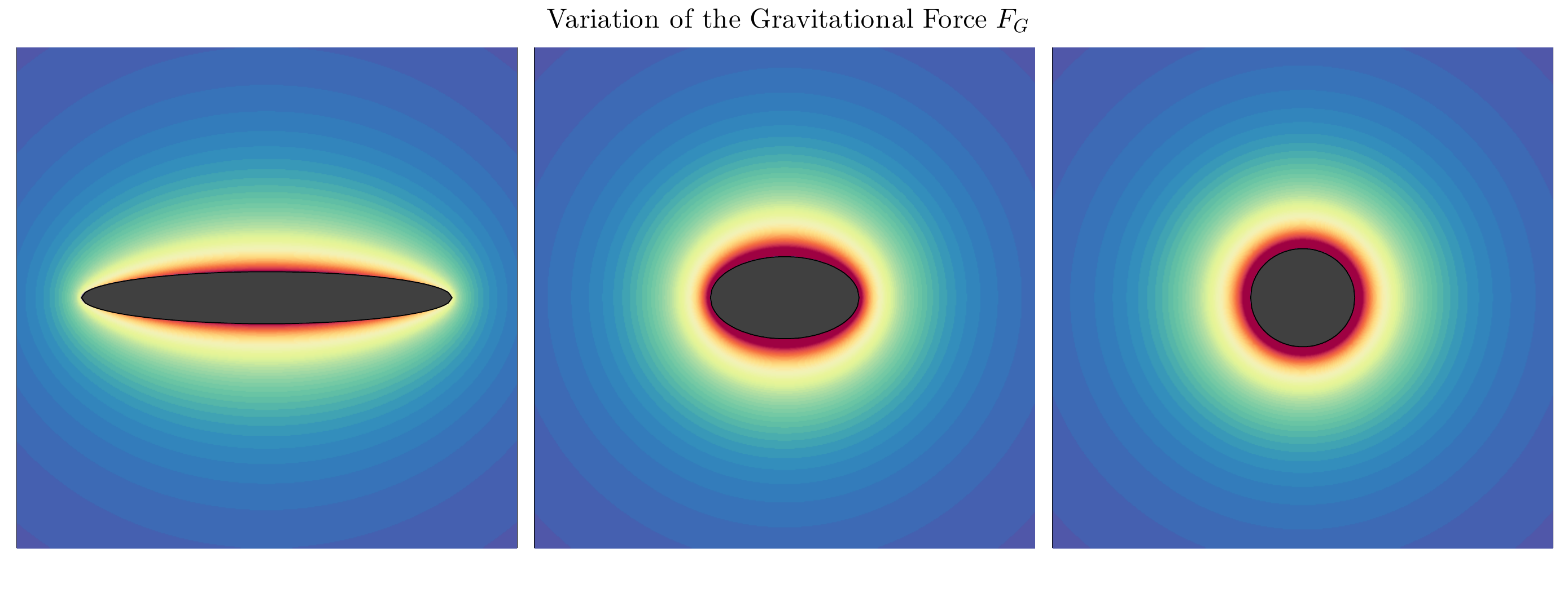} 
\caption{The shape dependence of the gravitational force.  This also represents  the shape dependence of the unscreened chameleon force around objects that do not have a thin shell.}\label{fig:Different_Ellipsoids_Grav}
\vspace{0.4cm}
\end{subfigure}
\begin{subfigure}{\textwidth}
\includegraphics[width=\textwidth]{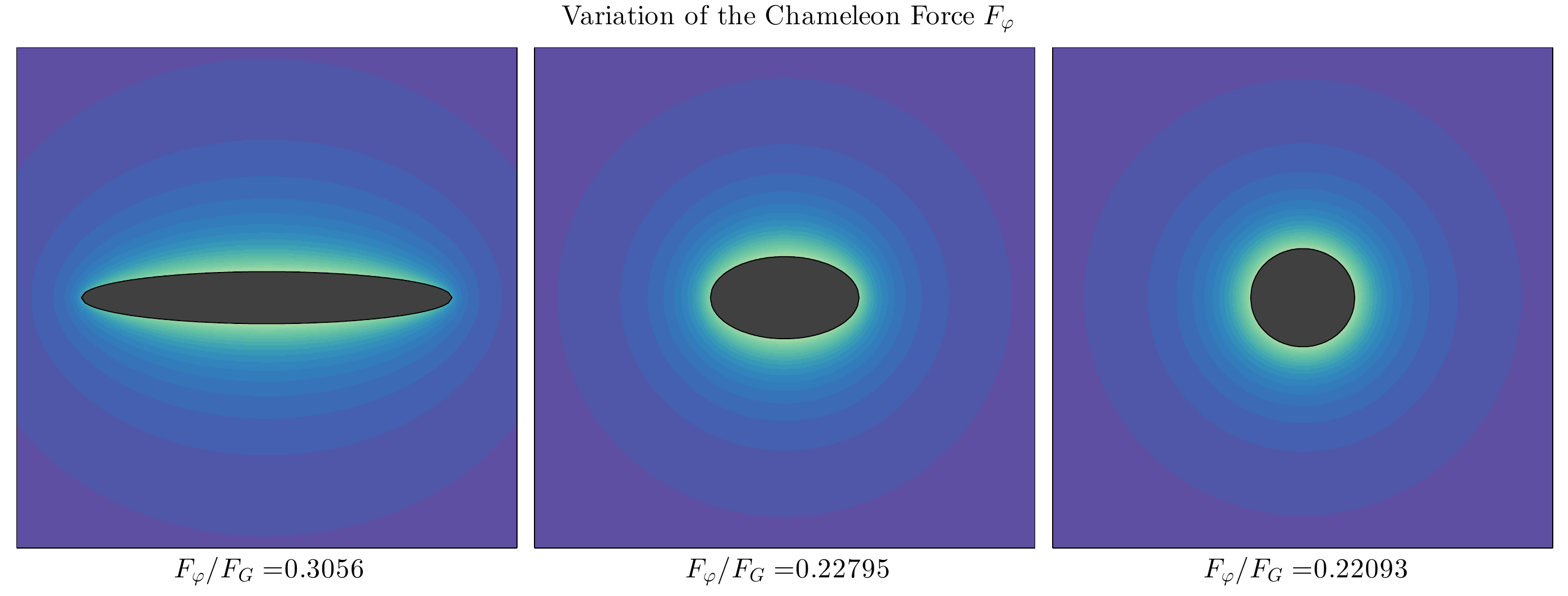}
\caption{The shape dependence of the chameleon force characteristic of objects for which a shell region has developed.} \label{fig:Different_Ellipsoids_Shell}
\end{subfigure}
\caption{Comparison of the shape dependence of the gravitational and chameleon force.  The mass of the source, represented by the black region, is the same for all plots.   From left to right the ellipsoids have $\xi_0 = 1.01$, $\xi_0 = 1.1$ and $\xi_0 = 5$ in both subfigures, corresponding to ellipticities $\sim 0.99$, 0.91 and 0.2 respectively.  The colours indicate the strength of the force, with red indicating regions of strongest force and blue indicating regions of weakest force.  The colour spectrum across these images has been normalised to the same limits to highlight the chameleons relative contribution. Increasing the ellipticity of the source can be seen to increase the ratio between the chameleon and the gravitational interactions. The parameter values  $\Lambda = 10^{-12}$ GeV and $M = 0.5M_{Pl}$ were chosen to generate these plots, but the shape dependence is independent of these choices.}
\label{fig:Different_Ellipsoids}
\end{figure}
Due to the nature of prolate spheroidal coordinates, variations in the ellipsoidal structure are most sensitive at low $\xi$, whilst the value of $\xi_0 = 5$ is visually almost indistinguishable from a perfect sphere. Accordingly, we take $\xi_0 = 5$ to represent the spherical limit. Figure \ref{fig:Different_Ellipsoids_Grav} shows the shape dependence of both the gravitational force and the chameleon force for a source exhibiting screening, due to the formation of a core region.  It is clear from figure \ref{fig:Different_Ellipsoids} that whilst the shape of the chameleon profile is the same as the shape of the gravitational profile for each ellipsoid, the ratio of the chameleon to gravitational force increases as the source becomes more ellipsoidal. This can be attributed to the ellipsoidal source favouring a comparatively smaller core component, resulting in less effective screening. Such behaviour is displayed in figure \ref{fig:Core_Scaling} which shows how the volume of the core scales with the compression of an object. 
As the internal chameleon structure responds to deformations by reducing the core region, there will be some threshold ellipticity where screening is deactivated, as the core vanishes. It follows that the onset of forming a shell is delayed as the source becomes more ellipsoidal, as can be seen from figure \ref{Sphere_Ellipse_Comparison}.  This figure shows the ratio of the chameleon and gravitational forces for different sized objects with different ellipticities.  For ease of comparison we take the specific choice $M=M_{PL}$ in order to generate this figure, but the shape dependence of the effect is independent of the choice of $M$.  When $F_{\varphi}/F_G=2$ there is no thin shell and the chameleon force is unscreened.  As the size of the source is increased, as parameterised by the effective radius $R_{\rm EFF}$, a thin shell develops and the chameleon force is suppressed.  Figure \ref{Sphere_Ellipse_Comparison} shows that this occurs for larger $R_{\rm EFF}$ when the source is more ellipsoidal, or equivalently has a smaller $\xi_0$.  This figure also demonstrates that the shape enhancement persists as the size of the source is increased. A conservative analysis of these results shows that the impact of deforming a sphere to an ellipsoid of $\xi_0 = 1.01$ enhances the chameleon force by as much as $40\%$. The difficulty in determining $\xi_{\rm core}$ analytically, makes it difficult to determine whether there is a maximum shape enhancement possible, or whether it continues to increase as the source is made more ellipsoidal.  However, as there is a limit to the ellipticity of an object that can be created, the behaviour of the shape enhancement as the ellipticity is increased much beyond the values considered here is not relevant for any realistic experimental scenario.

\begin{figure}[ht]
\begin{center}
\centerline{\includegraphics[width=\textwidth]{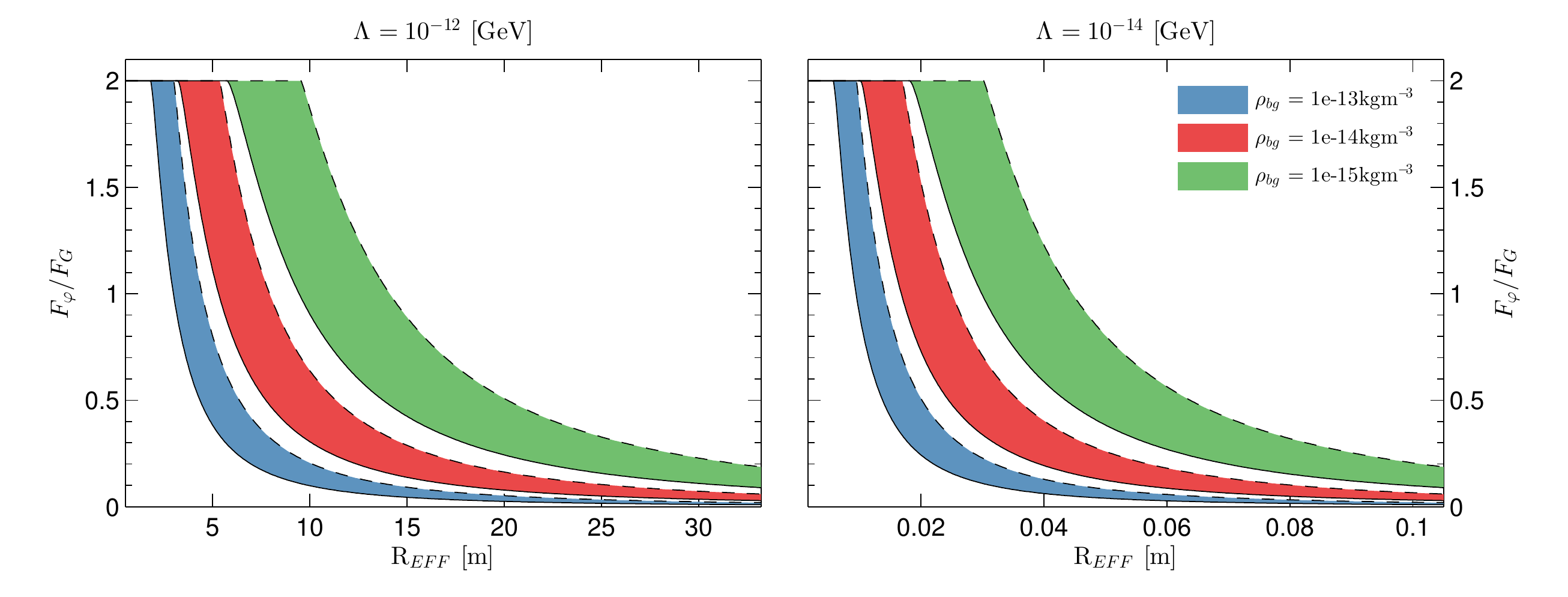}}  
\caption{The ratio of the chameleon to gravitational force as a function of the size of the source for different choices of $\Lambda$ and $\rho_{\rm bg}$.  
The lower edge of each band, drawn with a solid line, shows the force ratio for a spherical source.  The upper edge of each band, drawn with a dashed line, shows the force ratio for an ellipsoid with $\xi_0 = 1.001$. }
\label{Scalar_Grav_Ellipse_Density}
\end{center}
\end{figure}

Current experimental constraints on the chameleon from Casimir experiments, atomic spectroscopy and torsion balance tests of gravity 
\cite{Mota:2006fz,Gannouji:2010fc,Upadhye:2012qu,Brax:2010gp}  restrict the chameleon parameters to lie in the ranges $ \Lambda \leq 10^{-10}$ GeV and $10^4$ GeV $\leq M \leq M_{Pl}$. We restrict the values of $\Lambda$ that we study to the range $10^{-14} \mbox{ GeV} \leq \Lambda \leq 10^{-10} \mbox{ GeV}$ as we are most interested in chameleon theories with $\Lambda \sim 10^{-12} \mbox{ GeV}$ that can be connected to the dark energy scale.   Varying these parameters varies the size of object which forms a thin shell, as shown in figure \ref{Scalar_Grav_Ellipse_Density}.

It also shows how the onset of screening varies with the background density $\rho_{\rm bg}$.  
As the density of the environment is decreased the magnitude of the shape enhancement is seen to increase, as shown by the widths of the coloured bands.  Laboratory experiments searching for the chameleon can be made more sensitive if the background density is decreased, this increases the value of $\phi_{\rm bg}$ and makes the chameleon field harder to screen.  However creating increasingly more diffuse vacuums is a difficult task.  We have shown here that increasing the ellipticity of the source can increase the chameleon field strength in a similar manner to decreasing the density of the vacuum within which the experiment is performed.  Changing the shape of the source is an easier technical challenge than decreasing the density of the vacuum chamber and could therefore be used to increase the sensitivity of current proposals to detect the chameleon field, including that of \cite{CL}. Comparison between the two plots in figure \ref{Scalar_Grav_Ellipse_Density}  shows how varying $\Lambda$ changes the size of the source for which a thin shell first forms. This allows any laboratory experiment to increase the range of its sensitivity to $\Lambda$ by searching for the chameleon around both spherical and ellipsoidal sources. In principle, one could tune the size and shape of the chameleon source as to exploit this enhancement to maximise the detectability of a given point in the chameleon parameter space. As the value of $\Lambda$ is increased the difference between the chameleon field around an ellipsoidal and a spherical source decreases. 
\begin{figure}[ht]
\begin{center}
\centerline{\includegraphics[width=\textwidth]{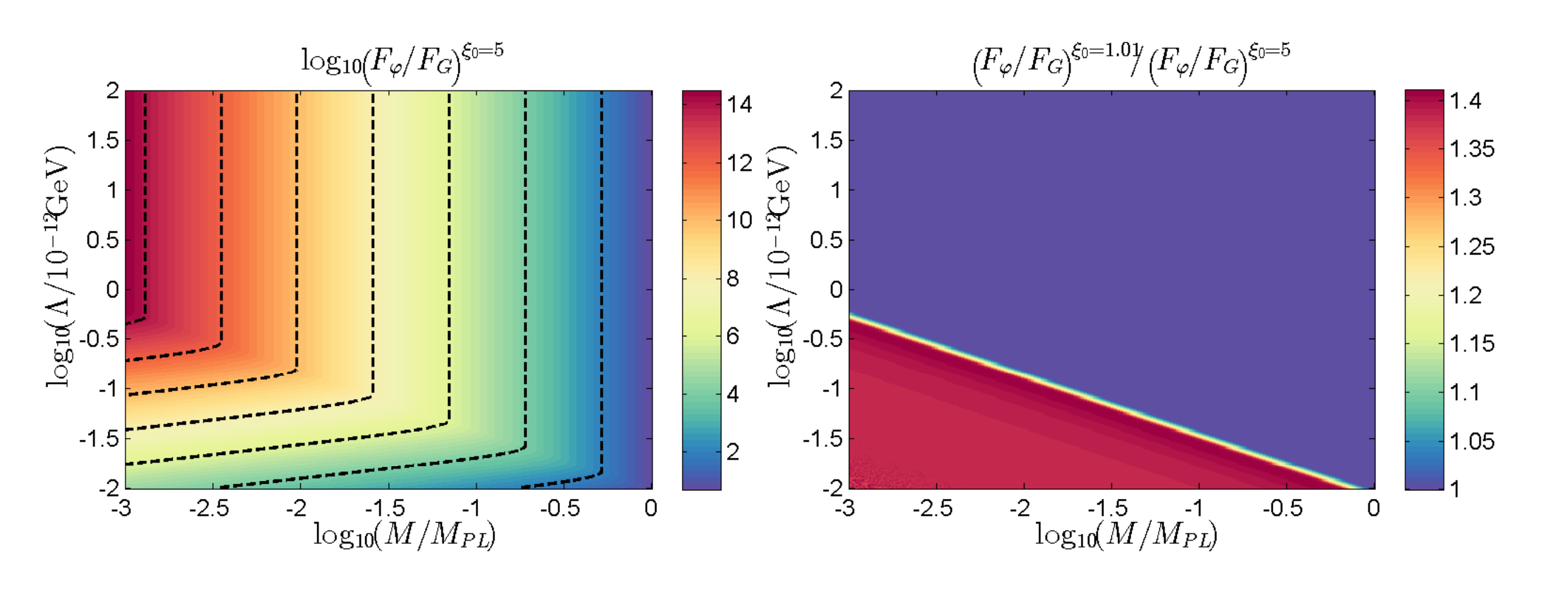}}  
\caption{Left: The ratio of the chameleon to gravitational forces as a function of the  parameters $\Lambda$ and $M$ for a spherical source object with $R_{\rm EFF} = 1$ cm.  The ambient and source densities take our reference values, $1$ gcm$^{-3}$ and $10^{-17}$ gcm$^{-3}$.  Right: The enhancement of  the chameleon force when an ellipsoid with $\xi_0 = 1.01$ is compared to a sphere.}
\label{fig:Vary_Parameters}
\end{center}
\end{figure}
The available chameleon parameter space is shown in figure \ref{fig:Vary_Parameters}.  The left hand plot shows the ratio of the chameleon to gravitational forces for a sphere in a laboratory scenario chosen to correspond to that of the experiment proposed in reference \cite{CL}.  The right hand plot shows the large region of parameter space in which the chameleon force can be enhanced by changing from a spherical to an ellipsoidal source. 
\section{Conclusions}
We have shown that increasing the ellipticity of a source, whilst keeping the density and mass of the source fixed, increases the ratio of the screened chameleon to gravitational forces. This is not because the chameleon force gets stronger but instead  the gravitational force suffers far more from the deformation. 
This behaviour is illustrated in figure \ref{fig:Different_Ellipsoids}. Thus when performing an experiment to search for a chameleon, replacing a spherical source with an ellipsoid of identical mass enhances the effects of the chameleon when compared to gravity thereby increasing the sensitivity of the experiment.  For realistic choices of parameters we find enhancements of the chameleon force by up to 40\%.  
We have also shown that working with a  sufficiently flattened disk  emulates the effects of  reducing the density of the laboratory vacuum by up to an order of magnitude. Thus, to increase chances of detection, changing the geometry of the source can be used as a cost effective alternative to  lowering the ambient density. \\
\\
This result shows that when considering the effects of fifth forces which posses screening mechanisms, the effects of the force can be underestimated if only spherical sources are considered, and that the strength of the force can vary significantly with the shape of the source.  Ellipsoidal sources can be studied analytically and so are a useful test bed in which to study these effects. However it would be interesting to pursue this further and determine the shape of the source for which the chameleon enhancement is largest.  This is a topic for future study. 
\section*{Acknowledgements}
We would like to thank Ed Hinds for very helpful discussions during the completion of this work. CB is supported by a Royal Society University Research Fellowship.  EJC is supported in part by the UK STFC. JS is supported by the Royal Society.
\appendix
\section{Spherical Prolate Coordinates}\label{A1}
The appropriate coordinate system to describe ellipsoidal geometries is prolate spheroidal coordinates. As described in the text, these are defined such that the major axis  of the ellipse lies along $z$ with foci at $ z = \pm a $. They relate to Cartesian co-ordinates via the following transformations:
\begin{equation}
\begin{split}
& x = a\sqrt{(\xi^2 - 1)(1 - \eta^2)}\> \cos(\phi)\;, \\
& y = a\sqrt{(\xi^2 - 1)(1 - \eta^2)}\> \sin(\phi)\;, \\
& z = a\xi\eta\;.
\label{eq:coords} 
\end{split}
\end{equation} 
The co-ordinate $\phi$ represents the conventional azimuthal angle appearing in spherical polar co-ordinates whereas $\xi$ and $\eta$ are analogous to the radial and polar angular components respectively, with $ +1 < \xi < \infty$, $-1 < \eta < +1$ and $ 0 < \phi < 2\pi$. This allows for the calculation of the associated scale factors 
\begin{equation}\label{scale_factors}
\begin{split}
& h_\xi  = \sqrt{\left(\frac{\partial x}{\partial \xi}\right)^2 + \left(\frac{\partial y}{\partial \xi}\right)^2 + \left(\frac{\partial z}{\partial \xi}\right)^2}  = a\>\sqrt{\frac{\xi^2 - \eta^2}{\xi^2 - 1}}\;, \\
& h_\eta = \sqrt{\left(\frac{\partial x}{\partial \eta}\right)^2 + \left(\frac{\partial y}{\partial \eta}\right)^2 + \left(\frac{\partial z}{\partial \eta}\right)^2}  = a\>\sqrt{\frac{\xi^2 - \eta^2}{1 - \eta^2} }\;,\\
& h_\phi = \sqrt{\left(\frac{\partial x}{\partial \phi}\right)^2 + \left(\frac{\partial y}{\partial \phi}\right)^2 + \left(\frac{\partial z}{\partial \phi}\right)^2}  = a\>\sqrt{(\xi^2 - 1)(1 -\eta^2)}\;,\\
\end{split}
\end{equation}
which can be used to determine the form of the Laplacian in prolate spheroidal coordinates:
\begin{align}\label{laplacian_transform}
\nabla^2\varphi &= \frac{1}{h_\xi h_\eta h_\phi} \sum_{i = 1}^{3}{\frac{\partial}{\partial\xi_i} \frac{h_\xi h_\eta h_{\phi}}{h_i ^2}\frac{\partial\varphi}{\partial\xi_i}}\;,  \\
&= \frac{1}{a^2(\xi^2 - \eta^2)}\left\{\frac{\partial}{\partial\xi}(\xi^2 - 1)\frac{\partial\varphi}{\partial\xi} + \frac{\partial}{\partial\eta}(1 - \eta^2)\frac{\partial\varphi}{\partial\eta} + \frac{\xi^2 - \eta^2}{(\xi^2 - 1)(1 - \eta^2)}\frac{\partial^2\varphi}{\partial\phi^2}\right\} \;.
\end{align}

\section{Computing the Massive Chameleon}\label{A2}
In this appendix we justify the assumption, made in the main body of the text, that the chameleon mass can be neglected. Beginning with the equation of motion 
\begin{equation}
\nabla^2 \varphi = m^2 (\varphi-\varphi_{\rm min}^{\rm (bg)})\;,
\end{equation}
we perform the field redefinition  $\psi = \varphi - \varphi_{\rm min}^{\rm (bg)}$ for simplicity,  and write the mass term as $ m^2 = - h^2/a^2$ in order to recover the Helmholtz equation in its canonical form, being careful to remember that $h$ is now imaginary. In spherical prolate coordinates the equation of motion we aim to solve is therefore:
\begin{equation}\label{Helmholtz}
\frac{1}{(\xi^2 - \eta^2)}\left\{\frac{\partial}{\partial\xi}(\xi^2 - 1)\frac{\partial\psi}{\partial\xi} + \frac{\partial}{\partial\eta}(1 - \eta^2)\frac{\partial\psi}{\partial\eta} + \frac{\xi^2 - \eta^2}{(\xi^2 - 1)(1 - \eta^2)}\frac{\partial^2\psi}{\partial\phi^2}\right\} = -h^2\psi\;.
\end{equation} 
\\
This has separable solutions of the form $ \psi = \tilde{\Phi}(\phi)\sum_{l=0}^{\infty}X_l(\xi)H_l(\eta) $ governed by the equations
\begin{align}
\frac{\partial^2\tilde{\Phi}(\phi)}{\partial\phi^2} + \kappa^2 \tilde{\Phi}(\phi) & = 0\;, \label{azimuthal} \\
\frac{\partial}{\partial\xi}(\xi^2 - 1)\frac{\partial X_l(\xi)}{\partial\xi} - \bigg(\frac{\kappa^2}{\xi^2 - 1} - h^2\xi^2 + \lambda_l\bigg)X_l(\xi) & = 0 \;,\label{Radial-EOM} \\
\frac{\partial}{\partial\eta}(1 - \eta^2)\frac{\partial H_l(\eta)}{\partial\eta} + \bigg(\frac{-\kappa^2}{1 - \eta^2} -h^2\eta^2 + \lambda_l\bigg)H_l(\eta) & = 0\;,  \label{Angular-EOM}
\end{align}
where $\kappa$ and $\lambda_l$ are separation constants, which we choose to be real. 
Equation \eqref{azimuthal} has trigonometric solutions of the form $\tilde{\Phi}(\phi) = \, ^{\cos}_{\sin}(\kappa\phi)$. The equations for $X_l(\xi)$ and $H_l(\eta)$ are identical, and only differ in the ranges of the dependent coordinates $\xi$ and $\eta$. The equation for $H_l(\eta)$ will prove easier to solve, and therefore we will address this first, the solution for $X_l(\xi)$ will then be determined from this by an integral transform. 
We are interested in solutions with azimuthal symmetry, and so we choose to set $\kappa=0$ implying that $\tilde{\Phi}(\phi) = \tilde{\Phi}_0 = {\rm const}$. 
With this choice of $\kappa$, equation (\ref{eta-equation})  resembles the Legendre equation
\begin{equation}\label{Legendre-Equation}
 (\eta^2 - 1)\frac{\partial^2 f_n}{\partial\eta^2} + 2\eta\frac{\partial f_n}{\partial\eta} - n(n+1)f_n = 0 \;,
\end{equation}
that has solutions $f_n = P_n, Q_n$, the Legendre functions of the first and second kind respectively. In fact as we will shortly see, the resemblance becomes exact in the massless case, when $h=0$ and $\lambda_l = l(l+1)$ where we have introduced $\lambda_l$ as a way of marking out the particular value of $l$ we are considering.  Given this, we will attempt to solve the equation of motion for $H_l(\eta)$ as a linear superposition of Legendre functions 
\begin{equation}\label{Heta-eqn}
H_l(\eta) = \sum_{n=0}^\infty A^{l}_n P_n (\eta),
\end{equation}
for constant  $A^l_n$. The reason we only consider $P_n (\eta)$ in equation (\ref{Heta-eqn}) is that the Legendre functions of the second kind, $Q_n (\eta)$ are only defined in the range $\eta >1$ and we are restricted here to $-1\leq \eta \leq 1$. 
Substituting equation (\ref{Heta-eqn})  into \eqref{Angular-EOM} (with $\kappa=0$) and utilising equation (\ref{Legendre-Equation}) to replace the derivative terms we obtain 
\begin{equation}\label{Subd-EOM}
\sum_{n=0}^\infty(h^2\eta^2 + n(n+1) - \lambda_l)A^{l}_n P_n= 0\;.
\end{equation} 
To determine the coefficients $A^{l}_n$ and the separation constants $\lambda_l$ we  make use of the recursion relation for Legendre Polynomials;
\begin{equation}
\eta^2 P_n = \frac{n(n-1)}{4n^2 - 1} P_{n-2} + \bigg[\frac{n^2}{4n^2 - 1} + \frac{(n+1)^2}{(2n+1)(2n + 3)}\bigg]P_n + \frac{(n+1)(n+2)}{(2n+1)(2n+3)}P_{n+2}\;,
\end{equation}
which when substituted into equation (\ref{Subd-EOM}) gives the following recursive relationship
\begin{equation}\label{Subd-EOM2}
\begin{split}
\sum_{n=0}^\infty  \bigg[  h^2\left\{\frac{n^2}{4n^2 - 1} + \frac{(n+1)^2}{(2n+1)(2n + 3)}\right\}&  + (n(n+1) - \lambda_l)\bigg]A^{l}_n P_n +  \\ 
&  h^2\frac{n(n-1)}{4n^2 - 1} P_{n-2}A^{l}_n + h^2\frac{(n+1)(n+2)}{(2n+1)(2n+3)}A^{l}_n P_{n+2} = 0 \;.
\end{split}
\end{equation}
As $P_{n<0}=0$ we choose to set $A^{l}_{n<0}=0$ and rearrange each of the individual summations in Equation (\ref{Subd-EOM2}) to obtain
\begin{equation}
\begin{split}
\sum_{n=0}^\infty \bigg[ h^2 \frac{(n+2)(n+1)}{(2n+3)(2n+5)} A^l_{n+2} + & \left\{\Big(\frac{n^2}{4n^2-1} + \frac{(n+1)^2}{(2n+1)(2n+3)}\Big)h^2 + n(n+1) - \lambda_l\right\}A^l_n +  \\
 & h^2 \frac{n(n-1)}{(2n-3)(2n-1)}A^l_{n-2}\bigg]P_n = 0\;.
\end{split}
\end{equation} 
As the Legendre polynomials are orthogonal over the interval (-1,1), this equation imposes a three-term recursion formula relating coefficients for all $n \in \mathbb{N}_0$
\begin{equation}\label{Recursion} 
\begin{split}
h^2 \frac{(n+2)(n+1)}{(2n+3)(2n+5)} A^l_{n+2} + & \left\{\Big(\frac{n^2}{4n^2-1} + \frac{(n+1)^2}{(2n+1)(2n+3)}\Big)h^2 + n(n+1) - \lambda_l\right\}A^l_n +  \\
 & h^2 \frac{n(n-1)}{(2n-3)(2n-1)}A^l_{n-2}= 0\;.
\end{split}
\end{equation}
Evidently, the above interconnects alternating coefficients and admits the trivial solution $A_l = 0 \>\>\forall\>\> l$ which applies for all $\lambda_l$.  Non-trivial solutions depend on either even or odd terms, and we will find that each of these series imposes different conditions on $\lambda_l$ and therefore both cannot  exist independently. The method of continued fractions may be applied to this recursion formula in order to generate a transcendental equation for $\lambda_l$. To continue it is convenient to present \eqref{Recursion} in a more compact fashion. Introducing three additional parameters $\alpha_n, \beta_n$ and $\gamma_n$ defined by
\begin{equation}\label{parameters-alpha-beta-gamma}
\begin{split}
&\alpha_n = h^2 \tilde{\alpha}_n~~~{\rm where}~\tilde{\alpha}_n = \frac{(n+2)(n+1)}{(2n+3)(2n+5)}\;, \\ 
&\beta_n = h^2 \tilde{\beta}_n + n(n+1) ~~~{\rm where}~\tilde{\beta}_n= \bigg[\frac{n^2}{4n^2 - 1} + \frac{(n+1)^2}{(2n+1)(2n+3)}\bigg]\;, \\ 
&\gamma_n= h^2\tilde{\gamma}_n~~~{\rm where}~\tilde{\gamma}_n= \frac{n(n-1)}{(2n-3)(2n-1)}\;,
\end{split}
\end{equation}
equation \eqref{Recursion}becomes
\begin{equation}\label{Recursion2}
\alpha_nA^l_{n+2} + (\beta_n - \lambda_l)A^l_n+ \gamma_nA^l_{n-2} = 0\;.
\end{equation}
Going further we define  $N^{l}_n = -\alpha_{n-2}\frac{A^{l}_n}{A^{l}_{n-2}}$ and $\delta_n = \gamma_n\alpha_{n-2}$ to arrive at the expression
\begin{equation}
N^l_n = \frac{\delta_n}{\beta_n - \lambda_l - N^l_{n+2}}\;, \hspace{1 in} n \geq 2\;.
\end{equation}
This allows us to construct an infinite continued fraction for the quantity $N^{l}_n$
\begin{equation}\label{Cont_Frac}
N^{l}_n = \cfrac{\delta_n}{\beta_n - \lambda_l - \cfrac{\delta_{n+2}}{\beta_{n+2} - \lambda_l - \cfrac{\delta_{n+4}}{\beta_{n+4} - \lambda_l - ...}}} \> \equiv \>  \polter{\delta_n}{\beta_n - \lambda_l} \> - \> \polter{\delta_{n+2}}{\beta_{n+2} - \lambda_l} \> - \> \polter{\delta_{n+4}}{\beta_{n+4} - \lambda_l} \> - \> ...
\end{equation}
where the second equality  employs the standard \textit{Pringsheim} notation as a more concise representation. Using  the recursion formula \eqref{Recursion2} to determine the leading terms in the series ($ n = 0$ and $ n = 1$) we find
\begin{equation}\label{Rec_S}
N^{l}_2 = \beta_0 - \lambda_l\;, \hspace{1 in} N^{l}_3 = \beta_1 - \lambda_l\;,
\end{equation} 
where
\begin{equation}\label{lambda_even}
0 = \beta_0 - \lambda_l - \polter{\delta_{2}}{\beta_{2} - \lambda_l} \> + \> \polter{\delta_{4}}{\beta_{4} - \lambda_l} \> + \> \polter{\delta_{6}}{\beta_{6} - \lambda_l} \> - \> ...
\end{equation}
A similar expression exists for the odd entries:
\begin{equation}\label{lambda_odd}
0 = \beta_1 - \lambda_l - \polter{\delta_{3}}{\beta_{3} - \lambda_l} \> + \> \polter{\delta_{5}}{\beta_{5} - \lambda_l} \> + \> \polter{\delta_{7}}{\beta_{7} - \lambda_l} \> - \> ...
\end{equation}
The roots of these two transcendental equations correspond to the various eigenvalues $\lambda_l$ for the even and odd expansion of the series solution, and can be solved to any order required. See for example references \cite{MF1} and \cite{SBP} for more details of the general case.
We will be only interested in the small mass limit ($|h|=ma \ll 1$) and  so we continue to present the solution for $H_l(\eta)$ in that case. 

\noindent We consider solutions for the coefficients $A_n^l$ and $\lambda_l$ in (\ref{Recursion2}) by expanding them in terms of series expansions in $h^2$ 
\begin{eqnarray}\label{pert-soln-hsq-anl}
A_n^l &=& \sum_{i=0}^{\infty} a^l_{n,i} h^{2i} \\
\lambda_l &=& \sum_{i=0}^{\infty} b_{l,i} h^{2i} \label{pert-soln-hsq-hl}
\end{eqnarray}
Substituting equations (\ref{pert-soln-hsq-anl}) and (\ref{pert-soln-hsq-hl}) into equation (\ref{Recursion2}) and using equation (\ref{parameters-alpha-beta-gamma}) we obtain 
\begin{equation}\label{Recursion3}
\sum_{i=0}^\infty \left[ \tilde{\alpha}_n a^l_{n+2,i}h^{2i+2} + \tilde{\gamma}_n  a^l_{n-2,i}h^{2i+2} + \tilde{\beta}_n  a^l_{n,i}h^{2i+2} + n(n+1)a^l_{n,i}h^{2i} - a^l_{n,i} \sum_{j=0}^\infty b_{l,j} h^{2i+2j} \right] =0 
\end{equation}
We now write equation (\ref{Recursion3}) order by order in the small parameter $h^2$ up to $O(h^4)$:
\begin{eqnarray}\label{h0-eqn}
h^0 &:& n(n+1)a^l_{n,0} - a^l_{n,0} b_{l,0} =0\\
h^2 &:& \tilde{\alpha}_n a^l_{n+2,0} + \tilde{\gamma}_n  a^l_{n-2,0} + \tilde{\beta}_n  a^l_{n,0} + n(n+1)a^l_{n,1} - a^l_{n,1} b_{l,0} - a^l_{n,0} b_{l,1} = 0 \label{h2-eqn} \\
h^4 &:& \tilde{\alpha}_n a^l_{n+2,1} + \tilde{\gamma}_n  a^l_{n-2,1} + \tilde{\beta}_n  a^l_{n,1} + n(n+1)a^l_{n,2} - a^l_{n,2} b_{l,0} - a^l_{n,1} b_{l,1}  - a^l_{n,0} b_{l,2}= 0 \label{h4-eqn}
\end{eqnarray}
The solution to equation (\ref{h0-eqn}) is 
\begin{equation}\label{soln-h0}
b_{l,0} =l(l+1),~~~~ a^l_{n,0}= a_0 \delta_{n,l}
\end{equation}
where $a_0$ is an unknown constant and $\delta_{n,l}$ is the Kronecker delta function. Substituting equation (\ref{soln-h0}) into equation (\ref{h2-eqn}) we obtain
\begin{equation}\label{h2-eqn1}
\tilde{\alpha}_n a_0 \delta_{n+2,l} + \tilde{\gamma}_n  a_0 \delta_{n-2,l} + \tilde{\beta}_n  a_0 \delta_{n,l} + n(n+1)a^l_{n,1} - a^l_{n,1} l(l+1) - a_0 \delta_{n,l} b_{l,1} = 0
\end{equation}
Prompted by the three Kronecker delta functions we consider the following cases, $n=l-2,\;n=l$ and $n=l+2$ in equation (\ref{h2-eqn1}). We then obtain
\begin{align}\label{al-2,1}
n=l-2\; &:& a^l_{l-2,1} &= {\tilde{\alpha}_{l-2} a_0 \over 2(2l-1)} \\
n=l\; &:& b_{l,1} &= \tilde{\beta}_l,\\
 & & a^l_{l,1} &= a^l_1~~{\rm an~unknown~constant}\label{bl,1} \\
n=l+2\; &:& a^l_{l+2,1} &= - {\tilde{\gamma}_{l+2} a_0 \over 4l+6} \label{al+2,1} \\
n \neq l-2,\;l,\;l+2\; &:& a^l_{n,1} &=0 \label{an,1}
\end{align} 
Substituting for $b_{l,0},\;a^l_{n,0}$ and $b_{l,1}$ in equation (\ref{h4-eqn}) we obtain
\begin{equation}
\tilde{\alpha}_n a^l_{n+2,1} + \tilde{\gamma}_n  a^l_{n-2,1} + \tilde{\beta}_n  a^l_{n,1} + n(n+1)a^l_{n,2} - a^l_{n,2} l(l+1) - a^l_{n,1} \tilde{\beta}_l  - a_0 \delta_{n,l} b_{l,2}= 0 \label{h4-eqn1}
\end{equation}
Following a similar approach as earlier we now consider the following cases, $n=l-4,\;n=l-2,\;n=l,\;n=l+2$ and $n=l+4$ in equation (\ref{h4-eqn1}) and obtain 
\begin{align}
\label{n=l-4}
n=l-4 \;&:& a^l_{l-4,2} &= - {\tilde{\alpha}_{l-4} \tilde{\alpha}_{l-2} a_0 \over 8(2l-1)(3-2l)} \\
n=l-2 \;&:& a^l_{l-2,2} &= {\tilde{\alpha}_{l-2} a^l_1 \over 2(2l-1)} + {(\tilde{\beta}_{l-2} - \tilde{\beta}_{l}) \tilde{\alpha}_{l-2} a_0 \over 4(2l-1)^2} \label{n=l-2} \\
n=l \;&:& b_{l,2} &= -{\tilde{\alpha}_l\tilde{\gamma}_{l+2} \over 2(2l+3)} +{\tilde{\alpha}_{l-2} \tilde{\gamma}_{l} \over 2(2l-1)},\\
 && a^l_{l,2} &= a^l_{2} = {\rm unknown~constant} \label{n=l}\\
n=l+2 \;&:& a^l_{l+2,2} &= {\tilde{\gamma}_{l+2} a^l_1 \over 2(2l+3)} + {(\tilde{\beta}_{l+2} - \tilde{\beta}_{l}) \tilde{\gamma}_{l+2} a_0 \over 4(2l+3)^2} \label{n=l+2} \\
n=l+4 \;&:& a^l_{l+4,2} &= {\tilde{\gamma}_{l+4} \tilde{\gamma}_{l+2} a_0 \over 8(2l+3)(2l+5)} \label{n=l+4}\\
n \neq l-4,\; l-2,\;l,\;l+2,\;l+4\; &:&  a^l_{n,2} &=0 \label{an,2}
\end{align}
We can begin to determine some of these coefficients. Consider the case of $l=0$. Using equation (\ref{parameters-alpha-beta-gamma}) in equations (\ref{soln-h0}), (\ref{al-2,1})-(\ref{an,1}) and  (\ref{n=l-4})-(\ref{n=l+4}) we obtain :
\begin{eqnarray}
b_{0,0} &=& 0;\;a^0_{n,0} = a_0 \delta_{n,0};\; b_{0,1}=\frac{1}{3}; \label{a0-coeff} \\
a^0_{0,1} &=& a_1 =~{\rm constant} ;\; a^0_{2,1} = -\frac{a_0}{9};\;a^0_{n,1} = 0~{\rm for}~n \neq 0, 2 \label{a1-coeff} \\
a^0_{0,2} &=& a^0_2 =~{\rm constant} ;\;b_{0,2}= - \frac{2}{135};\;a^0_{2,2} = -\frac{a_1}{9} + \frac{2 a_0}{567};\;a^0_{4,2}=\frac{a_0}{525} \label{a2-coeff}
\end{eqnarray}
It follows from equation (\ref{pert-soln-hsq-anl}) that 
\begin{eqnarray}
A^0_0 &=& a_0 + a_1 h^2 + a^0_2 h^4 + O(h^6) \label{A00} \\
A^0_2 &=& -\frac{a_0}{9}h^2 - \left(\frac{a_1}{9} - \frac{2 a_0}{567}\right)h^4 + O(h^6) \label{A02} \\
A^0_4 &=& \frac{a_0}{525}h^4 + O(h^6), \label{A04}
\end{eqnarray}
and from equation (\ref{pert-soln-hsq-hl}) that
\begin{equation}
\lambda_0 = \frac{1}{3} h^2 - \frac{2}{135} h^4 + O(h^6) \label{lambda0}.
\end{equation}
Now we have to determine the unknown constant coefficients $a_0,\;a_1$ and $a^0_2$. This can be done through correctly normalising our solutions, and so we impose that the angular solution $H_l(\eta)$ given in equation (\ref{Heta-eqn}) has the same normalisation factor as that of a single spherical harmonic $P_l(\eta)$. Recalling that 
\begin{equation}\label{PmPn-normalisation}
\int_{-1}^{1} P_m(\eta)P_k(\eta) d\eta =\frac{2}{(2m+1)} \delta_{m,k}
\end{equation} 
it follows that we require
\begin{equation}\label{normalisation-factor}
\int_{-1}^{1} (H_l(\eta))^2 d\eta = \frac{2}{(2l+1)}\;,
\end{equation}
Substituting equation (\ref{pert-soln-hsq-anl}) into equation (\ref{Heta-eqn}) and expanding to $O(h^4)$ we obtain
\begin{eqnarray}
H_l(\eta) &=& a_0 P_l + (a^l_{l-2,1} P_{l-2} + a^l_{l,1} P_{l} + a^l_{l+2,1} P_{l+2})h^2 \nonumber \\
&~& + (a^l_{l-4,2} P_{l-4} + a^l_{l-2,2} P_{l-2} + a^l_{l,2} P_{l}+ a^l_{l+2,2} P_{l+2} + a^l_{l+4,2} P_{l+4}) h^4 + O(h^6) \label{Hleta-pert}
\end{eqnarray}
Inserting into the normalisation condition equation (\ref{normalisation-factor}) and using equation (\ref{PmPn-normalisation}) we obtain
\begin{equation}\label{normalisation-factor1}
\int_{-1}^{1} (H_l(\eta))^2 d\eta = \frac{2}{2l+1} a_0^2 + \frac{4a_0 a_1}{2l+1} h^2 + \left[ \frac{4a_0 a^l_2}{2l+1} + \frac{2a_1^2}{2l+1} + \frac{2(a^l_{l-2,1})^2}{2l-3} + \frac{2(a^l_{l+2,1})^2}{2l+5}\right] h^4 = \frac{2}{(2l+1)}
\end{equation}
Solving equation (\ref{normalisation-factor1}) at each order in $h^2$ we obtain:
\begin{eqnarray}\label{h0-norm}
h^0 &:& a_0^2 =1 \to a_0 =1\\
h^2 &:& a_1 =0 \label{h2-norm}\\
h^4 &:&  a^l_2 = -\frac{1}{8} \left({l^2 (l-1)^2 \over (2l-3)(2l+1)(2l-1)^4} + {(l+2)^2(l+1)^2 \over (2l+5)(2l+1)(2l+3)^4}\right) \label{h4-norm}.
\end{eqnarray}
In particular for the case $l=0$, equation (\ref{h4-norm}) yields
\begin{equation}\label{a02}
a^0_2 = -\frac{1}{810}
\end{equation}
Substituting for $a_0,\; a_1$ and $a^0_2$ from equations (\ref{h0-norm}), (\ref{h2-norm}) and (\ref{a02}) into equations (\ref{A00})-(\ref{A04}) we obtain 
\begin{eqnarray}
A^0_0 &=& 1 -  \frac{1}{810} h^4 + O(h^6) \label{A00-1} \\
A^0_2 &=& -\frac{1}{9}h^2  + \frac{2}{567} h^4 + O(h^6) \label{A02-1} \\
A^0_4 &=& \frac{1}{525}h^4 + O(h^6). \label{A04-1}
\end{eqnarray}
Finally in equation (\ref{Heta-eqn}) we obtain an expression for the angular solution up to terms of order $h^4$:
\begin{eqnarray}
H_0(\eta) &=& A^0_0 P_0(\eta) + A^0_2 P_2(\eta) + A^0_4 P_4(\eta) \\
&=& \left(1 -  \frac{1}{810} h^4\right) P_0(\eta) - \left(\frac{1}{9}h^2  - \frac{2}{567} h^4\right)P_2(\eta) + \frac{1}{525}h^4 P_4(\eta) + O(h^6). 
\end{eqnarray}
It follows that the massless limit $h=0$ results in equality $H_0(\eta) = P_0(\eta)$. Moreover by considering $\lambda_l$ defined in equation (\ref{pert-soln-hsq-hl}) and using equation (\ref{soln-h0}) 
we see that 
\begin{equation}\label{lambdal-massless}
\lambda_l = l(l+1) + O(h^2)  
\end{equation}
which justifies the Legendre form of the equations we have used in the massless limit. This technique can of course be generalised to derive $H_l(\eta)$ for $l >0$, but we leave that as an exercise for the reader. Our purpose here was primarily to demonstrate how we could accommodate small mass corrections and we have demonstrated the technique to do that in this appendix.

\subsection{The Radial Solutions}
Although as can be seen by comparing equations (\ref{Radial-EOM}) and (\ref{Angular-EOM}), the radial component $X_l(\xi)$ of the field obeys the same differential equation as the angular component $H_l(\eta)$, it is defined over the coordinate range $1 \leq \xi \leq \infty$ rather than $-1 \leq \eta \leq 1$. Since the angular solution $H_l(\eta)$ is comprised of Legendre polynomials that are only defined over the interval $[-1,1 ]$ they are not applicable as solutions to the radial equation (\ref{Radial-EOM}).    Moreover, the method employed to derive the angular solution utilised a recursion relation that does not hold for Legendre functions of the second kind and so we cannot repeat this analysis to determine $X_l(\xi)$.  Nonetheless, a solution valid over the appropriate coordinate range can be found; a result presented in \cite{MF1} shows that, if a function $H_l(\eta)$ is a solution to a problem of the form \eqref{Angular-EOM}, then another solution, say $\chi^l(h\xi)$ may be obtained by performing an integral transform. The transform of the angular solution, in its general form $H_l(\eta) = \sum_{n=0}^\infty A^{l}_n P_n(\eta)$, corresponds to
\begin{equation}
\chi^l(h\xi) = \sum_{n=0}^{\infty}A_n^l\int_{-1}^{1}\>e^{ih\xi\eta}\>P_n(\eta)\>d\eta\;.
\end{equation}
This can be solved by 
employing the following relation between the Legendre polynomials and Bessel functions of the first kind \begin{equation}
J_{n+\frac{1}{2}}(h\xi) = (i^n\sqrt{2\pi})^{-1}\Big(\frac{h\xi}{2}\Big)^{1/2}\int_{-1}^{1}e^{ih\xi\eta} \> P_n(\eta)\>d\eta\;,
\end{equation}
and a second, linearly independent solution can be found using a related identity for Bessel functions of the second kind  $Y_{n+\frac{1}{2}}(h\xi)$ as they have the same derivative properties.  Together, the two independent solutions read
\begin{equation}\label{Both_Bessel_Sol}
\begin{split}
\chi^l_1(h\xi) = \Big(\frac{2\pi}{h\xi}\Big)^{1/2}\sum_{n=0}^{\infty}A_n^l\>i^n\>J_{n+\frac{1}{2}}(h\xi)\;, \\
\chi^l_2(h\xi) = \Big(\frac{2\pi}{h\xi}\Big)^{1/2}\sum_{n=0}^{\infty}A_n^l\>i^n\>Y_{n+\frac{1}{2}}(h\xi) \;. \\
\end{split}
\end{equation}
The general solution is therefore constructed by taking a linear superposition of the above
\begin{equation}\label{Radial_General}
X_l(\xi) = W^l\chi^l_1(h\xi) + V^l\chi^l_2(h\xi)\;,
\end{equation}
with the index $l$ being introduced to allow for the integration parameters $W^l,V^l$ to vary with different values of the separation constant, $\lambda_l $. Reintroducing the chameleon mass, $m=-i h/a$, we find the full solution 
\begin{equation}\label{Full-solution-with-mass}
\psi(\xi,\eta) =  \sum_{l= 0}^\infty\sum_{k=0}^\infty A_k^{l}P_k(\eta) \sum_{n=0}^{\infty}  \Big(\frac{2\pi}{ma\xi}\Big)^{1/2} A_n^{l}i^{(n+3/2)}\Big\{W^{l} J_{n+\frac{1}{2}}(ima\xi) + V^{l} Y_{n + \frac{1}{2}}(ima\xi)\Big\}\;.
\end{equation}
The constant value introduced by the normalisation of the polar solution $\tilde{\Phi}_0$ has been absorbed into the values of $W^l, V^l$.  We turn our attention to the radial $X_l(\xi)$ solutions which have to be handled more carefully as there are superficial divergences in the $m\to 0$ limit. Firstly, we transform the radial Bessel functions appearing in \eqref{Radial_General} into modified Bessel functions using:
\begin{equation}
\begin{split}
&J_\sigma(ix) = e^{\frac{\sigma i \pi}{2}} I_\sigma(x) \;,\\
&Y_\sigma(ix) = ie^{\frac{\sigma i \pi}{2}}I_\sigma(x) - \frac{2}{\pi}e^{-\frac{\sigma i \pi}{2}}K_\sigma(x)\;, 
\end{split}
\end{equation}
where $I_\sigma(x)$ and $K_\sigma(x)$ correspond to exponentially growing and decaying functions respectively. The solution \eqref{Radial_General} then becomes:
\begin{equation}\label{General_Bessel}
X_l(\xi) = -\bigg(\frac{2\pi}{ima\xi}\Big)^{1/2} \sum_{n=0}^{\infty} A^l_ni^n\Big((W^l + i V^l)
e^{i\pi/4} e^{\frac{i n \pi}{2}} I_{n+\frac{1}{2}}(ma\xi)  - \frac{2V^l }{\pi}e^{-i\pi/4} e^{-\frac{i n \pi}{2}}K_{n+\frac{1}{2}}(ma\xi)\bigg)\;.
\end{equation}
Regularity as $\xi \to \infty$ implies that we must have $W^l + i V^l = 0$ in order to cancel the divergence arising in the $I_{n+\frac{1}{2}}(ma\xi)$ term. 
$X_l(\xi)$ then simplifies to
\begin{equation}\label{General_Bessel-simple}
X_l(\xi) = \frac{2}{\pi} \bigg(\frac{2\pi}{ima\xi}\Big)^{1/2} e^{-i\pi/4} V^l \sum_{n=0}^{\infty} A^l_ni^n  e^{-\frac{i n \pi}{2}}K_{n+\frac{1}{2}}(ma\xi)\;.
\end{equation}
It proves convenient to express equation (\ref{General_Bessel-simple}) in terms of a \textit{Hankel Function} of the first kind which is related to the Bessel function $K_\nu(z)$ via
\begin{equation}
K_\nu (z) = \frac{i \pi}{2} e^{i\pi \nu/2} H^{(1)}_\nu (ze^{i\pi/2})~~~~[-\pi<{\rm arg} z \leq \pi/2],
\end{equation}
from which it follows that 
\begin{equation}\label{General_Hankel}
X_l(\xi) = - W^l \bigg(\frac{2\pi}{ima\xi}\Big)^{1/2}  \sum_{n=0}^{\infty} A^l_n i^n  H^{(1)}_{n+\frac{1}{2}}(ima\xi)\;.
\end{equation}
Now recalling $\psi = \varphi - \varphi_{\rm min}^{\rm (bg)}$, it follows from equation (\ref{Full-solution-with-mass}) using equation (\ref{General_Hankel}) that the complete solution for the massive system is
\begin{equation} 
\varphi(\xi,\eta,\phi) =   - \Big(\frac{2\pi}{ima\xi}\Big)^{1/2}\sum_{l=0}^\infty \sum_{k=0}^\infty W^l A_k^{l}P_k(\eta) \sum_{n=0}^{\infty} A_n^{l}i^n H^{(1)}_{n+\frac{1}{2}}(ima\xi) \>+\> \varphi_{\rm min}^{\rm (bg)}\;.
\end{equation}
We are interested in the small mass limit, or more precisely the limit $ma \ll 1$. This translates to a small argument for the Hankel function (of the first kind), which obeys
\begin{equation}
\lim_{z\to 0}~H^{(1)}_\nu(z) = - \frac{i}{\pi}\Gamma(\nu)\left(\frac{1}{2}z\right)^{-\nu} \;,
\end{equation}
where $\Gamma(\nu)$ is the Gamma function.  
It follows that when $ma \ll 1$, we have \\
\begin{equation}\label{Sol_Limit}
\varphi(\xi,\eta,\phi) = \frac{1}{\sqrt{\pi}} \sum_{l=0}^\infty \sum_{k=0}^\infty W^l A_k^{l}P_k(\eta) \sum_{n=0}^{\infty} A_n^{l}\>\Gamma\left(n + \frac{1}{2}\right)\> \left({2\over ma \xi}\right)^{n+1}\>+\> \varphi_{\rm min}^{\rm (bg)} 
\end{equation}
On the surface there doesn't appear to be a well defined limit as $ma \to 0$, as each successive term in the sum over $n$ appears to diverge more and more rapidly as $n$ increases. However this isn't the case as we now demonstrate. Recall $A^l_n$ is given as an expansion in $ma$ through equation (\ref{pert-soln-hsq-anl}). Inserting this in equation (\ref{Sol_Limit}) we obtain 
\begin{equation}\label{Sol_Limit1}
\varphi(\xi,\eta,\phi) = \frac{1}{\sqrt{\pi}} \sum_{l=0}^\infty \sum_{k=0}^\infty W^l A^l_k P_k(\eta) \sum_{n=0}^{\infty} \sum_{j=0}^{\infty} a^l_{n,j} (-1)^j (ma)^{2j-n-1} \>\Gamma\left(n + \frac{1}{2}\right)\> \left({2\over \xi}\right)^{n+1}\>+\> \varphi_{\rm min}^{\rm (bg)} 
\end{equation}
This still doesn't look particularly good, but there are conditions on the parameter $a^l_{n,j}$. Extending the analysis which led to equations (\ref{an,1}) and (\ref{an,2}), we see that all the coefficients $a^l_{n,j}=0$ unless $l-2j \leq n \leq l+2j$. In particular for the non-zero coefficients $a^l_{n,j}$ we have the bound $2j-n-1\geq -1-l$. This is the most divergent possible term we can have in equation (\ref{Sol_Limit1}). However, we can cancel it by recalling that the coefficients $W^l$ are so far undetermined. We are completely free to rewrite $W^l = \tilde{W}^l (ma)^{(l+1)}$, and with that we guarantee that there are no divergent contributions to the series expansion in equation (\ref{Sol_Limit1}). Indeed in the $ma \to 0$ limit only the term $2j-n-1 = -(l+1)$ survives and all the rest go to zero. Incorporating $\tilde{W}^l$ we have 
\begin{equation}\label{Sol_Limit2}
\varphi(\xi,\eta,\phi) = \frac{1}{\sqrt{\pi}} \sum_{l=0}^\infty \sum_{k=0}^\infty \tilde{W}^l A^l_k P_k(\eta) \sum_{n=0}^{\infty} \sum_{j=0}^{\infty} a^l_{n,j} (-1)^j (ma)^{2j-n+l} \>\Gamma\left(n + \frac{1}{2}\right)\> \left({2\over \xi}\right)^{n+1}\>+\> \varphi_{\rm min}^{\rm (bg)} 
\end{equation}
where we now see that the term $(ma)^{2j-n+l}$ does not diverge as $m\rightarrow 0$. To close this section, we demonstrate consistency of this solution by showing that the general $l = 0$ solution for the massive case reduces to the solution of the massless problem when setting $m = 0$. As a reminder, the (exterior) chameleon field equation of (\ref{eq:chamethin}) for a massless system reduces to the Laplace equation, where the solutions to this problem were presented when discussing the external solutions for gravity.  Therefore, the target is to recover the $l = 0$ entry of:
\begin{equation}\label{eq:Grav_Laplace}
\varphi(\xi, \eta, \phi) = \sum_{l =0}^{\infty} \varphi_l(\xi,\eta, \phi) = \sum_{l =0}^{\infty} W_l P_l(\eta)Q_l(\xi) 
\end{equation}
from the massive $l = 0$ result:
\begin{equation} 
\varphi_0(\xi,\eta,\phi) = \frac{1}{\sqrt{\pi}} \sum_{k=0}^\infty \tilde{W}^0 A^0_k P_k(\eta) \sum_{n=0}^{\infty} \sum_{j=0}^{\infty} a^0_{n,j} (-1)^j (ma)^{2j-n} \>\Gamma\left(n + \frac{1}{2}\right)\> \left({2\over \xi}\right)^{n+1}\>+\> \varphi_{\rm min}^{\rm (bg)} 
\end{equation}
with the subscript of zero signifying that the above corresponds to the $l = 0$ component of the complete solution.  Quoting a result from the preceding discussion, in the massless case the only surviving series coefficient of the form $A_n^l$ is $A_l^l$ i.e when $l = n$ (see (\ref{h0-eqn}) and \eqref{pert-soln-hsq-anl}). Further, inspection of \eqref{Sol_Limit2} for $l = 0$ shows that all but one of the radial terms inherit a mass dependence, with the exception being when $2j = n$. Collectively, these results can be used to eliminate two of the three summations and we can write:
\begin{equation}
\varphi_0(\xi,\eta,\phi) = \frac{1}{\sqrt{\pi}}\> \tilde{W}^0 \>P_0(\eta) \sum_{n=0}^{\infty}  a^0_{n,n/2} \> (-1)^{n/2} \>\Gamma\left(n + \frac{1}{2}\right)\> \left({2\over \xi}\right)^{n+1}\>+\> \varphi_{\rm min}^{\rm (bg)} 
\end{equation}
We have previously determined $a^0_{n,n/2}$ up to $n = 4$ \eqref{a2-coeff}\eqref{h0-norm} to be 
\begin{equation}
a^0_{0,0} = 1\;, \hspace{1 cm} a^0_{2,1} = -\frac{1}{9}\;, \hspace{1 cm} a^0_{4,2} = \frac{1}{525}\;.
\end{equation}
\\
Expanding the summation explicitly the first three terms of the Taylor series are identified to be:
\begin{equation}
\varphi_0(\xi,\eta,\phi) = \frac{\tilde{W}^0}{\sqrt{\pi}\xi} P_0(\eta) \bigg\{\Gamma\Big(\frac{1}{2}\Big) + \frac{4}{9}\>\Gamma\Big(\frac{5}{2}\Big)\frac{1}{\xi^2} + \frac{16}{525}\>\Gamma\Big(\frac{9}{2}\Big)\frac{1}{\xi^4} \>+\> \ldots\bigg\} +\> \varphi_{\rm min}^{\rm (bg)}\;,
\end{equation}
substituting the numerical values of the gamma functions gives
\begin{equation}
\begin{split}
\varphi_0(\xi, \eta, \phi) = \tilde{W}^0 \>P_0(\eta)\bigg\{\frac{1}{\xi} + \frac{1}{3}\frac{1}{\xi^3} + \frac{1}{5}\frac{1}{\xi^5} \>+\> \ldots\bigg\} +\> \varphi_{\rm min}^{\rm (bg)}\;,
\end{split}
\end{equation}
The terms appearing in brackets correspond to the series expansion of the Legendre function of second kind $Q_0(\xi)$. Therefore, the massless result for $l = 0$ is identified to be:
\begin{equation}
\varphi_0(\xi, \eta, \phi) = \tilde{W}^0 P_0(\eta) Q_0(\xi)\;,
\end{equation}
as recovered for the $l = 0$ case when solving the exterior Laplace equation \eqref{eq:Grav_Laplace}.
\section{Relating Ellipsoidal to Spherical coordinates}\label{A3}
In the large distance limit
$\xi \gg 1$ the relationship between ellipsoidal and spherical coordinates that can be inferred from   Equations (\ref{eq:coords}) becomes:

\begin{equation}
\begin{split}
a\xi\sqrt{1 - \eta^2} &= r\text{sin}(\theta)   \;,\\ 
a\xi\eta &= r\text{cos}(\theta)  \;,\\ 
\end{split}
\end{equation}
This simplifies to 
\begin{equation}
\eta = \text{cos}(\theta)\;, \hspace{1 cm} \xi = \frac{r}{a}\;.
\end{equation}
We have described the size of the ellipsoids in this work in terms of a reference sphere of the same volume.  
Equating the volumes of a sphere of radius $R$  with an ellipsoid of focal length $a$ and surface position $\xi_0$: 
\begin{equation}
\frac{4}{3}\pi R^3 = \int dV_{\rm ellipsoid} = \int_0^{2\pi}\int_{-1}^1\int_1^{\xi_0} h_{\xi}h_{\eta} h_{\phi}\>\> d\xi d\eta d\phi\;.
\end{equation}
Recalling the scale factors
\begin{equation}
h_{\xi} = a\sqrt{\frac{\xi^2 - \eta^2}{\xi^2 - 1}}\;, \hspace{0.5 cm} h_{\eta} = a\sqrt{\frac{\xi^2 - \eta^2}{1 - \eta^2}}\;, \hspace{0.5 cm} h_{\phi} = a \sqrt{(\xi^2 - 1)(1 - \eta^2)}\;,
\end{equation}
we find 
\begin{equation}
R^3 = a^3\xi_0\Big(\xi_0^2 - 1\Big)\;.
\label{eq:reff}
\end{equation}
This  assigns an effective radius to an ellipsoid defined by $\xi_0$.

\end{document}